\documentclass[aps,preprint,groupedaddress]{revtex4}
\usepackage[dvips]{graphicx}
\usepackage{amsmath}
\usepackage{graphicx}
\usepackage{amsfonts}
\usepackage{amssymb}
\usepackage{epsfig}
\usepackage{subfigure}
\usepackage[usenames]{color}

\begin{document}
\bibliographystyle{apsrev}

\title{\bf Holographic Magnetic Star}

\author{Piyabut Burikham$^{1,2}$\thanks{Email:piyabut@gmail.com, piyabut.b@chula.ac.th}\hspace{1mm} and Tossaporn Chullaphan$^{1}$\thanks{Email:chullaphan.t@gmail.com} \\
$^1${\small {\em Theoretical High-Energy Physics and Cosmology
Group,
Department of Physics,}}\\
{\small {\em Faculty of Science, Chulalongkorn University, Bangkok
10330, Thailand.}}\\
$^2$ {\small {\em  Thailand Center of Excellence in Physics, CHE,
Ministry of Education, Bangkok 10400, Thailand.}}\\}

\begin{abstract}
A warm fermionic AdS star under a homogeneous magnetic field is explored.  We obtain the relativistic Landau levels by using Dirac equation and use the Tolman-Oppenheimer-Volkoff~(TOV) equation to study the physical profiles of the star.  Bulk properties such as sound speed, adiabatic index, and entropy
density within the star are calculated analytically and numerically.  Bulk temperature increases the mass limit of the AdS star but external magnetic field has the opposite effect.  The results are partially interpreted in terms of the pre-thermalization process of the gauge matter at the AdS boundary after the mass injection.  The entropy density is found to demonstrate similar temperature dependence as the magnetic black brane in the AdS in certain limits regardless of the different nature of the bulk and Hawking temperatures.  Total entropy of the AdS star is also found to be an increasing function of the bulk temperature and a decreasing function of the magnetic field, similar behaviour to the mass limit.  Since both total entropy and mass limit are global quantities, they could provide some hints to the value of entropy and energy of the dual gauge matter before and during the thermalization.

\vspace{5mm}


\end{abstract}
\maketitle

\newpage

\section{Introduction}

Duality between gravity and gauge theory in spacetime with different dimensionalities has been discovered by Maldacena in 1998~\cite{Maldacena:1997re}.  The type-IIB string theory in AdS$_{5}\times S_{5}$ is conjectured to be dual to a gauge theory in four dimensional Minkowski spacetime ($M_{4}$) at the boundary of the AdS space.  The correspondence can be used as a complementary method to study the strongly coupled gauge theory in four dimensional Minkowski spacetime, a cousin of quantum chromodynamics, by avoiding the uncontrollable non-perturbative calculation via the application of weak-strong duality.  We can deal with this problem by alternatively performing calculations in the tractable weakly interacting string theory in five~(plus five compact dimensions which provide details that are not relevant here) dimensional Anti de Sitter space (AdS$_{5}$).  The duality is extended to a finite temperature situation by adding a horizon in the radial coordinate~\cite{wit}.  The string theory in an AdS space with black hole horizon in the radial direction is proposed to be dual to a gauge theory at finite temperature.  The duality is made quantitative in the sense that the Hawking temperature in the bulk theory corresponds to the temperature of the gauge theory on the boundary.  The AdS/CFT correspondence provides the first string-theoretic example of the underlying generic principle of the holographic duality~(i.e. the holographic principle).

The idea of holographic duality was originally proposed by 't Hooft~\cite{'tHooft:1993gx} in a generic quantum gravity situation involving a gravitational horizon.  The precise string theoretic version was given by Susskind~\cite{Susskind:1994vu}.  When an object falls into a black hole, it will be stretched, torn apart into bits and eventually the bits will be smeared out over the horizon.  Consequently, all of the bulk information is spread over the horizonal surface resulting in an effective boundary description of the bulk theory.  The bulk world is holographically encoded on the boundary.  Connection between AdS space and holography was further clarified by Witten~\cite{wit0} after discovery of the AdS/CFT correspondence.

Given an AdS space, the weakly-coupled bulk gravity theory corresponds to a strongly-coupled boundary gauge theory.  Adding a black hole to the AdS space, the dual gauge theory on the boundary becomes thermal with the temperature equal to the corresponding Hawking-Page temperature of the background~\cite{hp}.  It is thus interesting to investigate the intermediate situation where there exists a massive object before gravitational collapse into a black hole in the AdS space and search for the dual description in the gauge theory side.  It is argued in Ref.~\cite{deBoer:2009wk,Arsiwalla:2010bt} that the degenerate fermions in the AdS correspond to the composite multitrace operator  constructed from product of single trace operators in the large central charge limit on the boundary.  It is not unreasonable to think of this ``free" fermionic operator as the conformal cousin of a QCD nucleon such as neutron and proton.  These ``free fermions", however, still interact with each other by the colour-singlet interaction of order $1/N$ assuming negligible in the large $N$ limit.  The colour-singlet~(glueball) exchange on the boundary corresponds holographically to the gravitational interaction in the bulk.  While gravity pulls the bulk mass together causing the gravitational collapse, the colour-singlet interaction should be responsible for the deconfinement phase transition of the injected mass in the dual picture.

Arguably, the gravitational collapse of the star in the AdS would correspond to a thermalization process of the dual gauge matter on the boundary~\cite{dkk,ssz,ls,cy,bm,bbb,bbbc}.  Consideration of the mass limit of the fermionic star in the AdS bulk could reveal certain details of the pre-thermalization process in the dual gauge picture.  The mass limit of the AdS star corresponds to the minimum amount of injected mass required in order for the boundary gauge matter to start the thermalization process~(since the bulk gravitational collapse starts when injected mass exceeds the mass limit).  Specifically, it is also interesting to ask what the dual object of the bulk temperature is on the boundary world before black hole formation?  Should it correspond to some parameter characterizing the superheated phase of gauge matter before the start of the thermalization?  Moreover, what is the exact nature of the colour-singlet~(glueball exchange) interaction responsible for the deconfinement of the dual gauge matter into the thermalized deconfined plasma~(which is the dual picture of the gravitational collapse caused by gravity)?

The heavy-ion collision experiments at RHIC and CERN's LHC~(Large Hadron Collider) smash two charged ions at extreme energies, producing dense and hot nuclear matter with properties of the strongly coupled plasma.  In the vicinity of the collision point, the induced magnetic field could be enormous~\cite{kmw}.  Understanding the physics of dense hot nuclear plasma under such circumstances requires nonperturbative treatments of the strong interaction and the holographic method is one option.  One holographic dual of the magnetized nuclear matter at finite temperature is proposed to be a magnetized black brane in the AdS space~\cite{hk}.  It was found that the entropy density of the magnetized brane in the AdS obeys the third law of thermodynamics with entropy $S\sim T$~(temperature) for small temperature.

In this article, we consider a fermionic star in the holographic AdS$_{5}$ background in the presence of external magnetic field at finite bulk temperature.  The mass limit and other properties of the star is studied with respect to the changes in the magnetic field and bulk temperature.  Even though there is no complete understanding of the dual description in the gauge theory side of this situation, we argue certain aspects of the duality.  In Section II, the Tolman-Oppenheimer-Volkoff~(TOV) equation~\cite{Tolman:1939jz,Oppenheimer:1939ne} in the background AdS$_{5}$ is calculated starting from the general dimensionality.  The energy levels of the bulk charged fermions in the presence of the magnetic field are calculated in the flat space approximation.  The pressure and density of the bulk fermions at finite field and temperature are subsequently derived.  Section III presents analytic and numerical results for each case of finite temperature and field.  The mass limits depend crucially on the field and bulk temperature.  The mass-radius relations for each case are discussed in Section IV.    The bulk adiabatic index and sound speed of the fermions inside the AdS star are discussed in Section V.  The entropy density and total entropy in the bulk are also computed.  Section VI investigates the dependence of mass limit on the AdS radius.  Section VII contains further discussions and summary of our results.

\section{Holographic Star under External Magnetic Field}

The study of the magnetized star in the AdS space consists of two main calculations.  First, the pressure and energy density need to be calculated for the system of charged fermions in the magnetic field at arbitrary temperature.  The star will be assumed electrically neutral and we will focus only on the effect of magnetic field to the charged particles.  At zero temperature, the energy states of the charged fermions in the magnetic field are separated naturally into Landau levels.  The partition function in the macrocanonical ensemble of these energy levels will provide the generic expression for the pressure and energy density of the fermionic system at finite temperature.  The pressure and energy density are subsequently used in the equation of state required by the TOV equation in the 5-dimensional AdS spacetime. Even though we will focus on interpreting the results of the bulk AdS star in terms of the dual gauge theory, the calculations in the bulk picture are self-consistent and satisfactorily describe a real magnetized fermionic star in the 5-dimensional AdS spacetime.

\subsection{The Equations of Hydrostatic Equilibrium for a Spherical Symmetric Star in $d$ dimensions}

In order to study the behaviour of a degenerate star in $d$-dimensional AdS spacetime, we derive the spherical symmetric TOV equation in $d$ dimensions as given in Appendix~\ref{appa}.  In the presence of external magnetic field, the pressure of the fermionic matter in the star is actually anisotropic due to the quantization of the energy levels.  However, in the classical limit where the momentum in the direction of the magnetic field is much larger than the square root of the magnetic field, $<p_{z}^{2}>/m^{2}c^{2}\gg 2Be\hbar/m^{2}c^{3}$, the pressure becomes isotropic~\cite{cc1,cc2} and the spherical symmetric TOV equation is applicable.  The resulting equations of motion describing the AdS star in the spherical symmetric approximation are $T(r)=T_{0}\mu(r)/\mu_{0}$ for the temperature $T(r)$, and
\begin{eqnarray}
M'\left(r\right) &=& \frac{2V_{d-2}}{\left(d-2\right)}\rho\left(r\right)r^{d-2}, \label{EoM of M} \\
\mu'\left(r\right) &=& \mu\left(r\right)\left(\frac{B'(r)}{B(r)} - \frac{V_{d-2}C_{d-1}}{\left(d-2\right)}\left(\rho\left(r\right)c^{2} + P_{r}\left(r\right)\right)rB^{2}\left(r\right)\right), \label{EoM of mu}   \end{eqnarray}
where $B(r)=(1 - \frac{MC_{d-1}}{r^{d-3}} + \frac{r^{2}}{l^{2}})^{-1/2}, l$ is the AdS radius, $V_{d-2}$ is the area of $S^{d-2}$ and $C_{d-1}=\frac{16\pi G}{\left(d - 2\right)V_{d-2}c^{4}}$.  To solve the equations of motion, we need the equation of state or the explicit expression of $P(r), \rho(r)$ in terms of the chemical potential $\mu(r)$.  Standard evaluation of the partition function requires the layout of energy states of the fermionic system which can be obtained in the following subsection.

\subsection{Relativistic Landau Energy Level in 5 dimensions}

We now solve the Dirac equation to find the relativistic energy level of a charged fermion in the presence of external magnetic field in the 5 dimensional spacetime.  As an approximation, we will ignore the effect of curvature on the energy levels of the fermions.  The effects of gravity and the AdS curvature will be considered only through the Einstein equations stated in the previous subsection.  Starting from the Dirac equation in flat space
    \begin{eqnarray}
        i\hbar\gamma^{\mu}\partial_{\mu}\psi - mc\psi = 0, \label{Dirac equation}
    \end{eqnarray}
where $m$ is the mass of the fermion.  The gamma matrices are chosen to be in the Dirac representation as the following
    \begin{eqnarray}
        \gamma^{0} =
    \begin{pmatrix}
        \mathbf{1} & 0  \\
        0 & -\mathbf{1}
    \end{pmatrix}, ~~~~~~
        \vec{\gamma} =
    \begin{pmatrix}
        0 & \vec{\sigma} \\
        -\vec{\sigma} & 0
    \end{pmatrix},
    \end{eqnarray}
where $\mathbf{1}$ and $\vec{\sigma}$ are $2\times2$ identity matrix and Pauli matrices respectively.  We will consider only the positive energy solution since we are interested in the particle not the antiparticle.  The positive energy solution $\psi\left(x\right) = u\left(p\right)e^{-ipx} = u\left(p\right)e^{-iEt+i\vec{p}\cdot\vec{x}}$ satisfies the equation $\left(\gamma^{\mu}p_{\mu}-m\right)u\left(p\right) = 0$.  Let $\hbar = 1$ and consider a particle in an external magnetic field, the effect of the magnetic field can be taken into account by adding the field momentum, $p_{\mu} \rightarrow p_{\mu} - qA_{\mu}$.  We will choose the magnetic field to point in the $z$ direction and uniformly distributed over the entire $x,y,z$  space.  The equation of motion of the fermion in 5 dimensional space becomes
    \begin{eqnarray}
        \{p^{2}_{x} + p^{2}_{y} + p^{2}_{z} + p^{2}_{w} - 2qBxp_{y} + q^{2}B^{2}x^{2} - qB\sigma_{z}\}\phi = \left(E^{2}-m^{2}c^{4}\right)\phi. \label{Find E-AdS eq}
    \end{eqnarray}
The momentum component in the extra dimension is represented by $p_{w}$ corresponding to the coordinate $w$.  We have assumed the solution in the form $\phi = e^{i\left(p_{y}y+p_{z}z+p_{w}w\right)}f\left(x\right)$ and neglect the effect of the AdS curvature to the momentum component $p_{w}$. This is a good approximation as long as the AdS radius of curvature is large compared to the wavelength of the bulk fermions.

The energy condition from the equation of motion is given by
    \begin{eqnarray}
        E^{2}_{n} = m^{2}c^{4} + p^{2}_{z}c^{2} + p^{2}_{w}c^{2} + \left(2n-\nu+1\right)2mc^{2}\mu_{B}B. ~ \left(n = 0, 1, 2, \ldots ~~, \nu = \pm1\right) \nonumber
    \end{eqnarray}
If we let $j = n - \frac{\nu}{2}$, then we have
    \begin{subequations}
    \begin{align}
        E^{2}_{j} &= m^{2}c^{4} + p^{2}_{z}c^{2} + p^{2}_{w}c^{2} + \left(j+\frac{1}{2}\right)4mc^{2}\mu_{B}B, \label{Landau energy level 1} \\
                  &= m^{2}c^{4} + p^{2}_{n}c^{2} + \left(j+\frac{1}{2}\right)4mc^{2}\mu_{B}B. ~~~~~~ \left(p^{2}_{n} = p^{2}_{z} + p^{2}_{w}\right) \label{Landau energy level 2}
    \end{align}
    \end{subequations}
From equation (\ref{Landau energy level 1}) and (\ref{Landau energy level 2}), energy is quantized in the $x-y$ plane and contains certain degeneracy of states, \textit{i.e.}, there are several states with the same one-particle energy. The number of states $g_{j}$ of a discrete energy level $j$ is
\begin{eqnarray}
        g_{j} &=& \frac{g_{s}}{h^{2}}\int dp_{x}dp_{y}dxdy = \frac{g_{s}L_{x}L_{y}}{h^{2}}2\pi\int_{p_{j}}^{p_{j+1}}pdp = \frac{g_{s}\pi L_{x}L_{y}}{h^{2}}\left(p^{2}_{j+1} - p^{2}_{j}\right), \nonumber \\ &=& \frac{g_{s}\pi L_{x}L_{y}}{h^{2}}\left(4m\mu_{B}B\right). ~~~~ \left( \because p^{2}_{j}c^{2} = \left(p^{2}_{x} + p^{2}_{y}\right)c^{2} = 4jmc^{2}\mu_{B}B\right)
\end{eqnarray}
where $g_{s}$($= 2s + 1$) is a spin degeneracy independent of $j$.  The degeneracy is proportional to the field and vanishes for $B \rightarrow 0$.  The discrete energies from the degrees of freedom of the plane perpendicular to the magnetic field is called the Landau levels, characterizing the statistical properties of the fermionic system.  Extension to finite temperature situation can be done by considering the corresponding partition function.

\subsection{Pressure and Energy Density under Magnetic Field at Finite Temperature}

Thermodynamical pressure and energy density of the magnetized fermion gas can be calculated from the grand canonical partition function given by
\begin{eqnarray}
        \ln Z &=& \frac{1}{h^{2}}\int_{-\infty}^{\infty}dp_{z}dp_{AdS}dzdx_{AdS} \sum_{j=0}^{\infty}g_{j}\ln\left(1+e^{-\frac{\left(E_{j}-\mu\right)}{k_{B}T}}\right), \nonumber \\ &=& \frac{g_{s}L_{x}L_{y}L_{z}L_{AdS}}{h^{4}}\left(4\pi m\mu_{B}B\right)\int_{-\infty}^{\infty}dp_{z}dp_{AdS} \sum_{j=0}^{\infty}\ln\left(1+e^{-\frac{\left(E_{j}-\mu\right)}{k_{B}T}}\right), \nonumber \\ &=& \left(\frac{4g_{s}\pi m\mu_{B}BV}{h^{4}}\right)\left(2\pi\right)\int_{0}^{\infty}p_{n}dp_{n} \sum_{j=0}^{\infty}\ln\left(1+e^{-\frac{\left(E_{j}-\mu\right)}{k_{B}T}}\right). \label{parf}
\end{eqnarray}
Use the Euler-Maclaurin formula~(see Appendix \ref{appb}) and certain tricks of integration, we finally have the pressure in the asymptotically approximated form
\begin{eqnarray}
        P & = & \frac{k_{B}T}{V}\ln Z = \frac{k_{B}T}{V}\left(\ln Z_{0} + \ln Z_{B}\right), \nonumber \\
        &\simeq & \left(\frac{g_{s}\pi^{2}}{2h^{4}}\right) \Bigg[\int_{mc^{2}}^{\mu}\left(\frac{\epsilon^{2}}{c^{2}} - m^{2}c^{2}\right)^{2}d\epsilon - k_{B}T\int_{0}^{\frac{\mu - mc^{2}}{k_{B}T}}\frac{\left(\frac{\left(\mu - k_{B}Ty\right)^{2}}{c^{2}} - m^{2}c^{2}\right)^{2}}{e^{y} + 1}dy  \nonumber \\
        && +~k_{B}T\int_{0}^{\infty}\frac{\left(\frac{\left(\mu + k_{B}Ty\right)^{2}}{c^{2}} - m^{2}c^{2}\right)^{2}}{e^{y} + 1}dy\Bigg] - \left(\frac{2\pi^{2}m^{2}\mu_{B}^{2}B^{2}}{3h^{4}}\right)\Bigg[\left(\mu - mc^{2}\right)  \nonumber \\
        && -~k_{B}T\int_{0}^{\frac{\mu - mc^{2}}{k_{B}T}}\frac{dy}{e^{y} + 1} + k_{B}T\int_{0}^{\infty}\frac{dy}{e^{y} + 1}\Bigg]. \label{Pressure in AdS space under external magnetic field at finite temperature}
\end{eqnarray}
Likewise the energy density is given by
\begin{eqnarray}
       U & = & \frac{g_{s}}{h^{2}}\int_{-\infty}^{\infty}dp_{z}dp_{AdS}dzdx_{AdS}\sum_{j=0}^{\infty}g_{j}\frac{E_{j}}{z^{-1}e^{\frac{E_{j}}{k_{B}T}} + 1}, \nonumber \\ &=& \frac{g_{s}L_{x}L_{y}L_{z}L_{AdS}}{h^{4}}\left(4\pi m\mu_{B}B\right)\int_{-\infty}^{\infty}dp_{z}dp_{AdS}\sum_{j=0}^{\infty}\frac{E_{j}}{z^{-1}e^{\frac{E_{j}}{k_{B}T}} + 1}, \nonumber \\
       & = & \left(\frac{8g_{s}\pi^{2}m\mu_{B}BV}{h^{4}}\right)\int_{0}^{\infty}p_{n}dp_{n}\sum_{j=0}^{\infty}\frac{E_{j}} {z^{-1}e^{\frac{E_{j}}{k_{B}T}} + 1}.
\end{eqnarray}
Again, use the Euler-Maclaurin formula and tricks of integration, so that $\rho c^{2} = \frac{U}{V} = \frac{U_{0} + U_{B}}{V}$ becomes
\begin{eqnarray}
        \rho c^{2} & \simeq & \left(\frac{2g_{s}\pi^{2}}{h^{4}c^{2}}\right)\Bigg[ \int_{mc^{2}}^{\mu}\epsilon^{2}\left(\frac{\epsilon^{2}}{c^{2}} - m^{2}c^{2}\right)d\epsilon - k_{B}T\int_{0}^{\frac{\mu - mc^{2}}{k_{B}T}}\frac{\left(\mu - k_{B}Ty\right)^{2}\left(\frac{\left(\mu - k_{B}Ty\right)^{2}}{c^{2}} - m^{2}c^{2}\right)}{\left(e^{y} + 1\right)}dy  \nonumber \\
        && + k_{B}T\int_{0}^{\infty}\frac{\left(\mu + k_{B}Ty\right)^{2}\left(\frac{\left(\mu + k_{B}Ty\right)^{2}}{c^{2}} - m^{2}c^{2}\right)}{\left(e^{y} + 1\right)}dy \Bigg] + \left(\frac{2g_{s}\pi^{2}m^{2}\mu_{B}^{2}B^{2}}{3h^{4}}\right)\Bigg[ (\mu - mc^{2}) \nonumber \\
        && - k_{B}T\int_{0}^{\frac{\mu - mc^{2}}{k_{B}T}} \frac{dy}{\left(e^{y} + 1\right)} + k_{B}T\int_{0}^{\infty} \frac{dy}{\left(e^{y} + 1\right)} - \int_{mc^{2}}^{\mu}\frac{\epsilon e^{\frac{\epsilon-\mu}{k_{B}T}}}{k_{B}T}d\epsilon - 2\int_{0}^{\frac{\mu - mc^{2}}{k_{B}T}}\frac{\left( \mu - k_{B}T y \right)e^{-y}}{\left( e^{y} + 1\right)}dy  \nonumber \\
        && +\int_{0}^{\frac{\mu - mc^{2}}{k_{B}T}}\frac{\left(\mu - k_{B}Ty\right)e^{-y}}{\left(e^{y} + 1\right)^{2}}dy + \int_{0}^{\infty}\frac{\left(\mu + k_{B}Ty\right)e^{y}}{\left(e^{y} + 1\right)^{2}}dy \Bigg]. \label{Energy density in AdS space under external magnetic field at finite temperature}
\end{eqnarray}
Both expressions for the pressure and energy density are in the remarkable form with the dependence on $B$ separated out in simple quadratic functions.  The integrations can be cast into logarithmic and polylogarithmic functions depending only on the temperature~(and not the field) as are shown in the next section.

\section{Numerical Results}

In this section, the equations of motion, Eqn.~(\ref{EoM of M}),(\ref{EoM of mu}) will be solved numerically.  To emphasize effects of both temperature and external magnetic field, the physical properties of the degenerate star in the AdS$_{5}$ under the influence of both temperature and external magnetic field are investigated by dividing into 4 cases; 1.) $T=0, B=0$, 2.) $B=0, T>0$, 3.) $B>0, T=0$, and 4.) $B,T>0$.
Before going into the details of each case, we integrate equations (\ref{Pressure in AdS space under external magnetic field at finite temperature}) and (\ref{Energy density in AdS space under external magnetic field at finite temperature}) to obtain
    \begin{eqnarray}
        P & = & \left(\frac{g_{s}\pi^{2}}{30c^{4}h^{4}}\right) \Bigg[ 3\mu(r)^{5} - 10m^{2}c^{4}\mu(r)^{3} + 15m^{4}c^{8}\mu(r) - 8m^{5}c^{10} - 10k_{B}^{2}T^{2}m^{2}c^{4}\pi^{2}\mu(r) \nonumber \\
        &&  +~7k_{B}^{4}T^{4}\pi^{4}\mu(r) + 10k_{B}^{2}T^{2}\pi^{2}\mu(r)^{3} - 120k_{B}^{3}T^{3}m^{2}c^{4}Li_{3}\left(-e^{\frac{mc^{2}-\mu(r)}{k_{B}T}}\right)  \nonumber \\
        &&  +~360k_{B}^{4}T^{4}mc^{2}Li_{4}\left(-e^{\frac{mc^{2}-\mu(r)}{k_{B}T}}\right) - 360k_{B}^{5}T^{5}Li_{5}\left(-e^{\frac{mc^{2}-\mu(r)}{k_{B}T}}\right)  \nonumber \\
        &&  -~20k_{B}Tm^{2}c^{4}\mu_{B}^{2}B^{2}\ln\left(1 + e^{\frac{\mu(r)-mc^{2}}{k_{B}T}}\right) \Bigg], \label{Pressure in Simulation}
    \end{eqnarray}

    \begin{eqnarray}
        \rho c^{2} & = & \left(\frac{g_{s}2\pi^{2}}{15c^{4}h^{4}}\right)
        \Bigg[ 3\mu(r)^{5} - 5m^{2}c^{4}\mu(r)^{3} + 2m^{5}c^{10} - 5k_{B}^{2}T^{2}m^{2}c^{4}\pi^{2}\mu(r) + 7k_{B}^{4}T^{4}\pi^{4}\mu(r)  \nonumber \\
        &&  +10k_{B}^{2}T^{2}\pi^{2}\mu(r)^{3} + 30k_{B}^{2}T^{2}m^{3}c^{6}Li_{2}\left(-e^{\frac{mc^{2}-\mu(r)}{k_{B}T}}\right) - 150k_{B}^{3}T^{3}m^{2}c^{4}Li_{3}\left(-e^{\frac{mc^{2}-\mu(r)}{k_{B}T}}\right)  \nonumber \\
        &&  +360k_{B}^{4}T^{4}mc^{2}Li_{4} \left( -e^{\frac{mc^{2}-\mu(r)}{k_{B}T}}\right) - 360k_{B}^{5}T^{5}Li_{5} \left( -e^{\frac{mc^{2}-\mu(r)}{k_{B}T}}\right) \Bigg] + \left(\frac{4m^{2}\pi^{2}\mu_{B}^{2}B^{2}}{3h^{4}}\right) \nonumber \\
        && \Bigg[\frac{mc^{2}}{1 + e^{\frac{\mu(r)-mc^{2}}{k_{B}T}}} - \mu(r) - k_{B}T\ln\left(1 + e^{\frac{mc^{2}-\mu(r)}{k_{B}T}}\right) + k_{B}T\ln\left(1 + e^{\frac{\mu(r)-mc^{2}}{k_{B}T}}\right) \Bigg], \label{Energy density in Simulation}
    \end{eqnarray}
where $Li_{s}(z) = \sum_{k=1}^{\infty}\frac{z^{k}}{k^{s}}$ is a polylogarithm function. For numerical analysis, we set $G_{5} = Gl$, $G = c = \hbar = k_{B} = \mu_{B} = l = 1$, $m = 0.1$.  We can transform the numerical results to the SI unit by using the table of dimensional translation given in Appendix~\ref{appc}.  The coupled equations of motion between mass and chemical potential (Eqn. (\ref{EoM of M}), (\ref{EoM of mu})) are numerically solved to find the chemical potential and the accumulated mass within the star.  The density and pressure profiles can be subsequently obtained.  The boundary conditions at the center of star are chosen to be $M(r = 0) = 0$ and $\mu(r = 0) = e \simeq 2.718281828$ for every case.

\subsection{Case I, zero temperature and zero magnetic field}

This is the condition of degenerate star in AdS$_{5}$ considered in Ref.~\cite{deBoer:2009wk}.  The fermions degenerate into the lowest possible energy states filling the energy levels up until the Fermi energy in 5 dimensions.  In this limit, the pressure and the energy density, Eqn. (\ref{Pressure in Simulation}), (\ref{Energy density in Simulation}), reduce to
    \begin{subequations}
    \begin{align}
        P =& \left(\frac{g_{s}\pi^{2}}{30c^{4}h^{4}}\right)\left(3\mu(r)^{5} - 10m^{2}c^{4}\mu(r)^{3} + 15m^{4}c^{8}\mu(r) - 8m^{5}c^{10}\right), \label{Pressure in Simulation at T=0, B=0} \\
        \rho c^{2} =& \left(\frac{g_{s}2\pi^{2}}{15c^{4}h^{4}}\right)\left(3\mu(r)^{5} - 5m^{2}c^{4}\mu(r)^{3} + 2m^{5}c^{10}\right). \label{Energy density in Simulation at T=0, B=0}
    \end{align}
    \end{subequations}
First, the surface of the star can be defined at the radial distance, $R$, where the pressure becomes zero.  Apparently from Eqn.~(\ref{Pressure in Simulation at T=0, B=0}), the pressure is zero when $\mu(r=R) = mc^{2}$.  On the other hand, from Eqn.~(\ref{Energy density in Simulation at T=0, B=0}), the density vanishes when $\mu/mc^{2}=-1.3848, 1$.  Therefore in this case, both the pressure and energy density become zero at the radius $R$ where $\mu(R)=mc^{2}$.

The accumulated mass, the chemical potential, the density and the pressure distribution of the star versus the radius are presented in Figure~\ref{T=0,B=0MMuvsr} and \ref{T=0,B=0RhoPvsr}.  Relations between the total mass and the central chemical potential/density of the degenerate star are shown in Figure~\ref{T=0,B=0MasslimitMu}.
    \begin{figure}[h]
        \centering
        \subfigure[]{\includegraphics[width=0.45\textwidth]{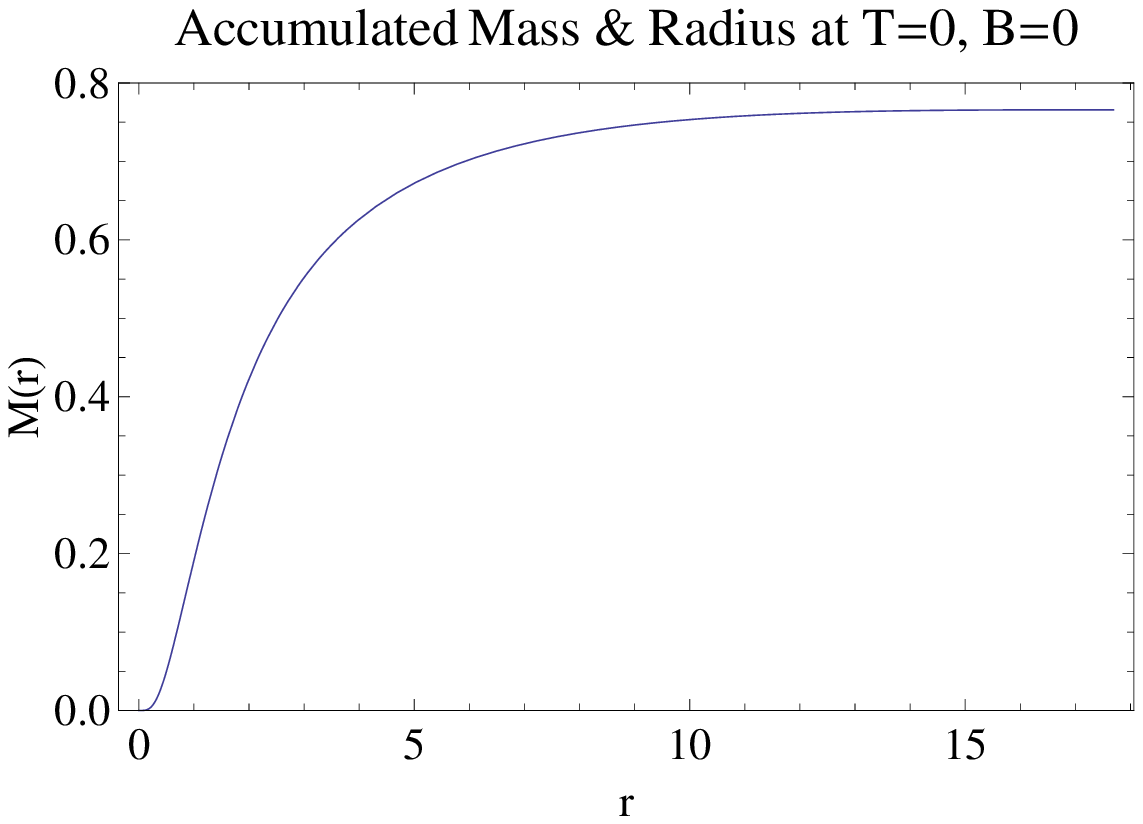}}\hfill
        \subfigure[]{\includegraphics[width=0.45\textwidth]{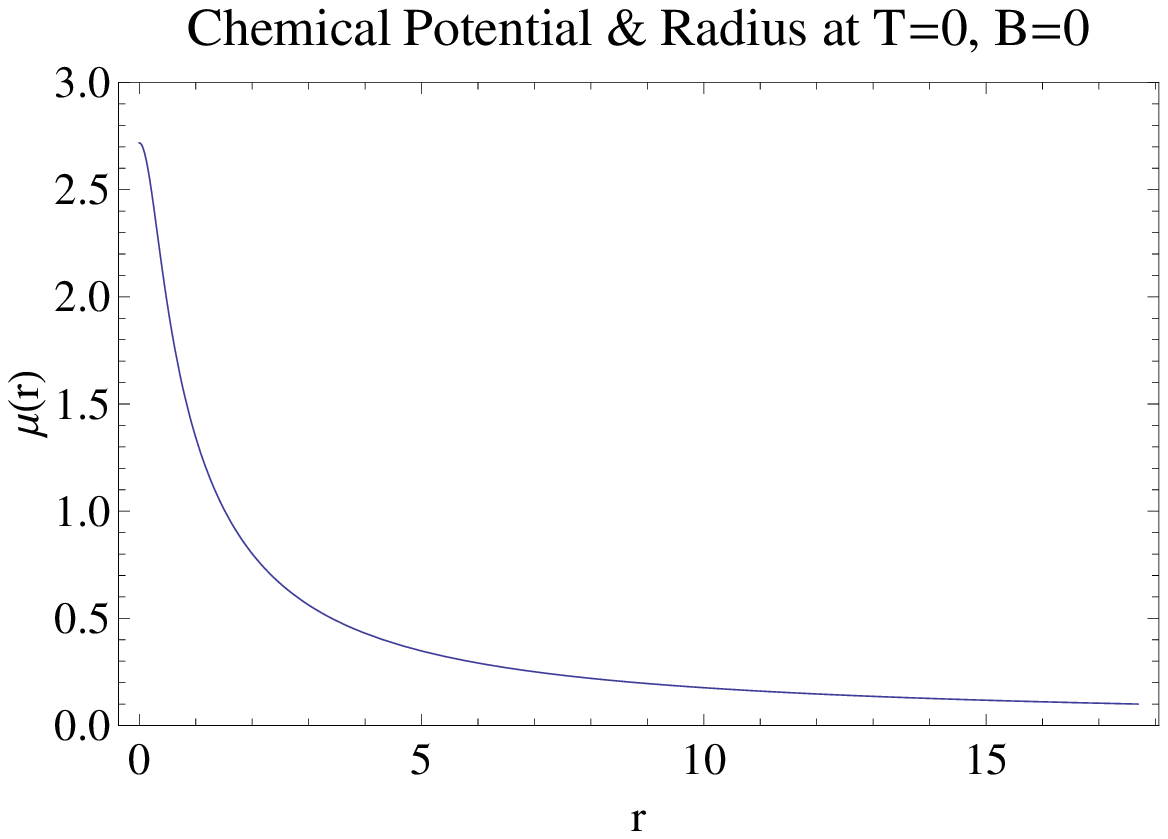}}\\
        \caption{The accumulated mass(a) and the chemical potential(b) distribution in the degenerate star at $T = 0$, $B = 0$} \label{T=0,B=0MMuvsr}
        \subfigure[]{\includegraphics[width=0.45\textwidth]{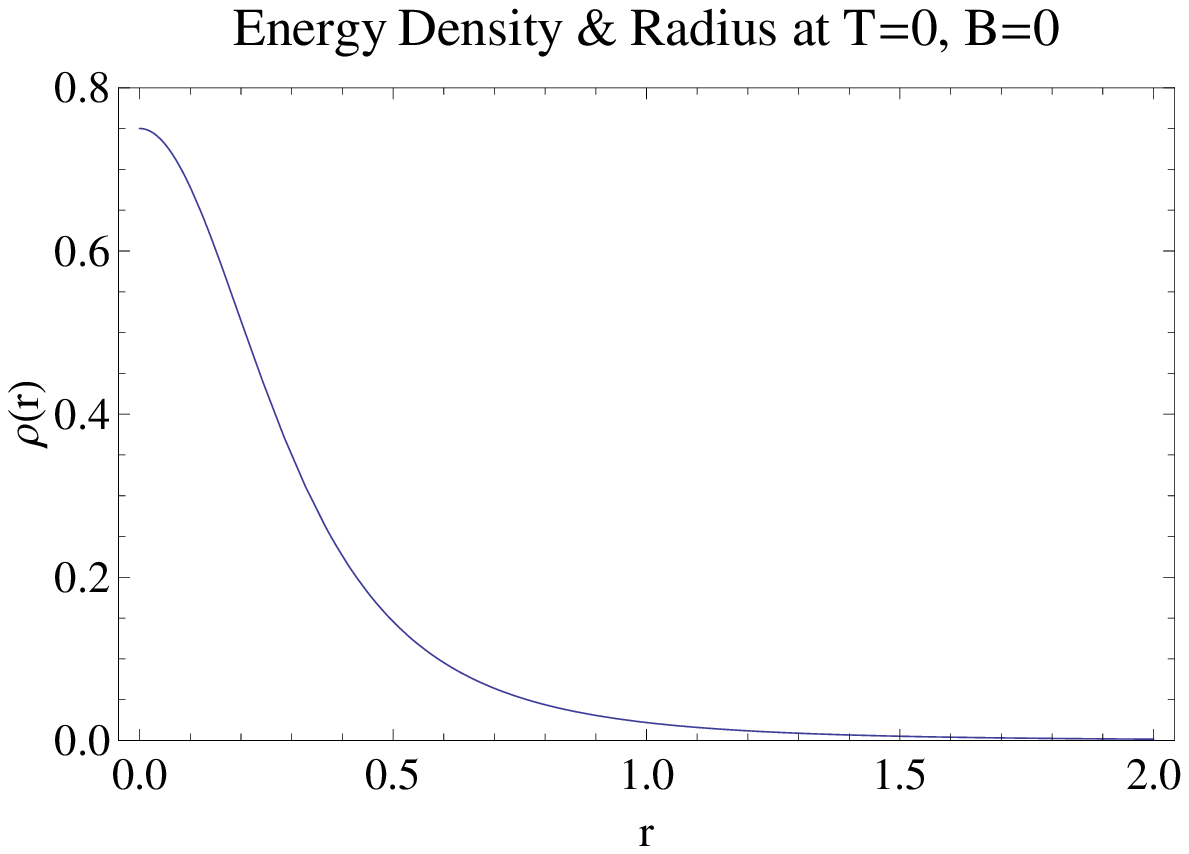}}\hfill
        \subfigure[]{\includegraphics[width=0.45\textwidth]{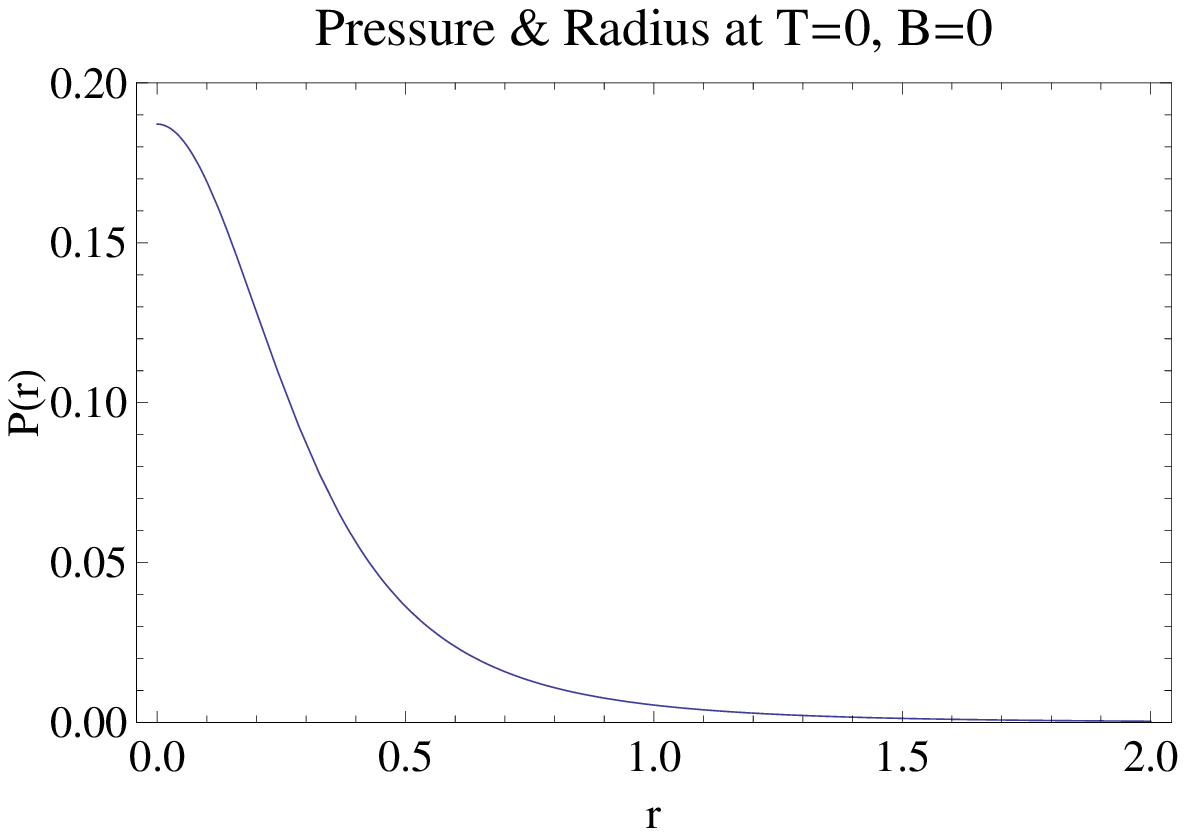}}\\
        \caption{The density(a) and the pressure(b) distribution in the degenerate star at $T = 0$, $B = 0$} \label{T=0,B=0RhoPvsr}
    \end{figure}
    \begin{figure}[h]
        \centering
        \subfigure[]{\includegraphics[width=0.45\textwidth]{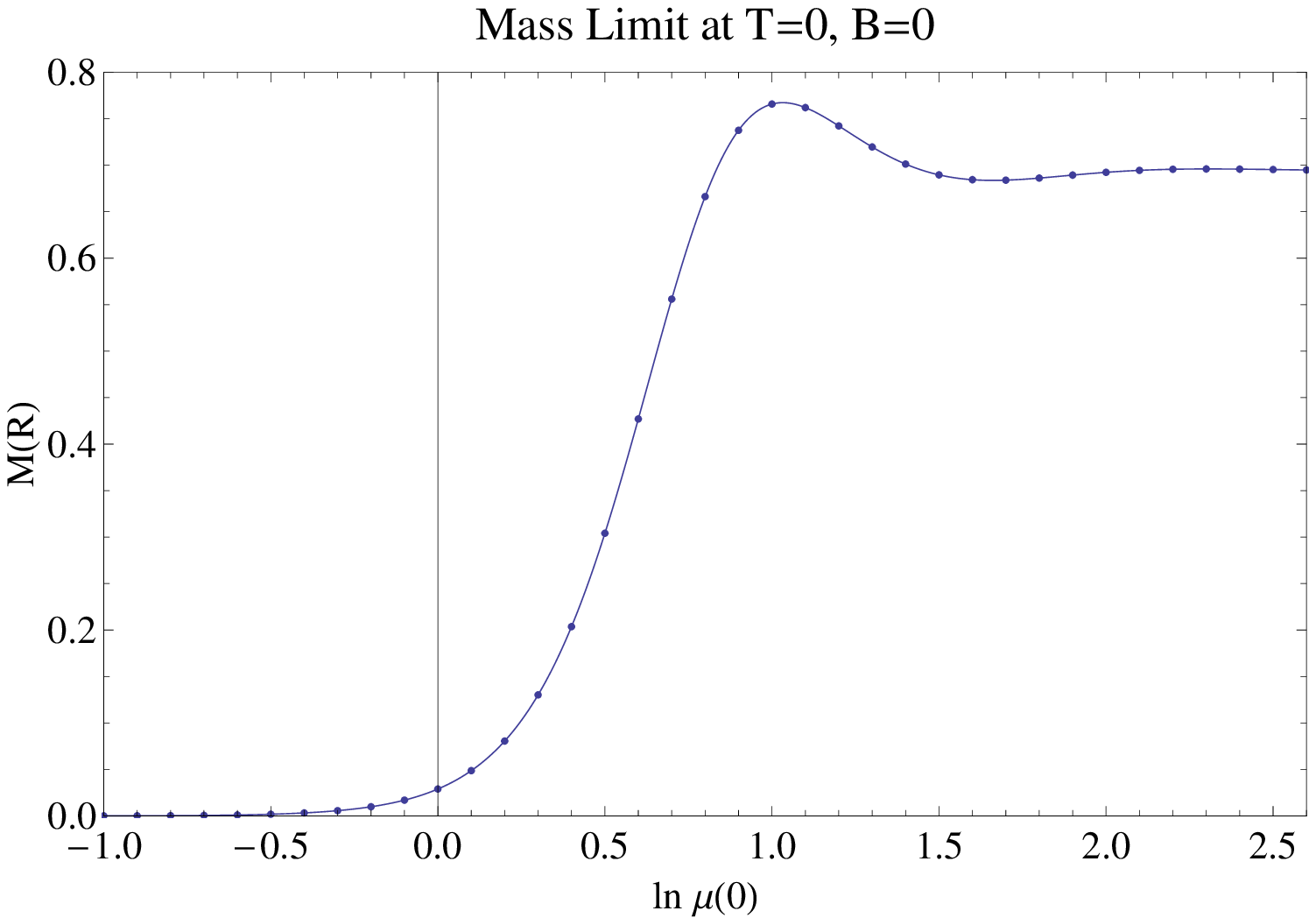}}\hfill
        \subfigure[]{\includegraphics[width=0.45\textwidth]{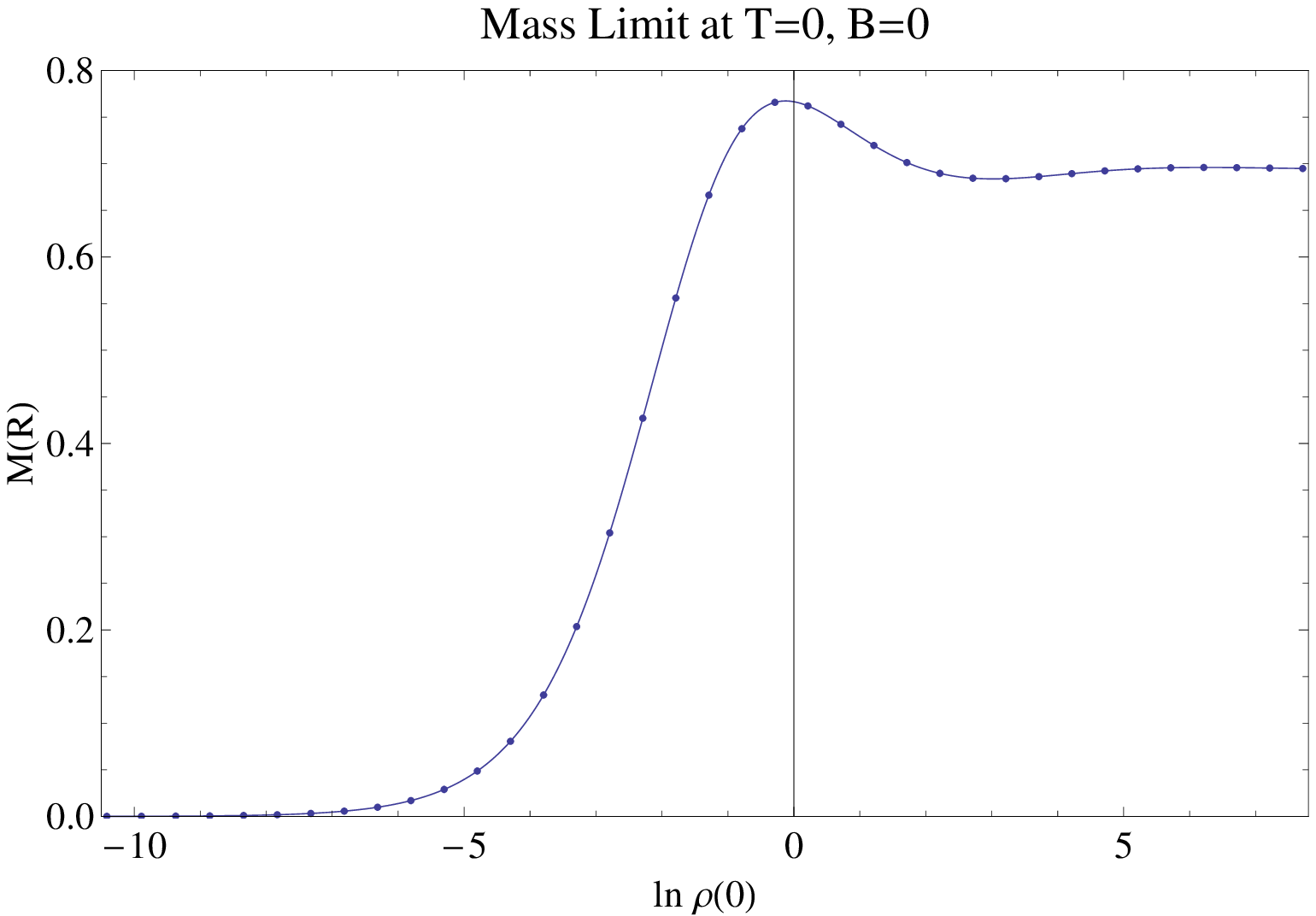}}\\
        \caption{The relation between mass and central chemical potential/density (in logarithmic scale) of the degenerate star at $T = 0$, $B = 0$} \label{T=0,B=0MasslimitMu}
    \end{figure}

From the numerical solution, the edge of the degenerate star is at $r = 17.6922$ where the pressure drops to zero. In Figure~\ref{T=0,B=0MMuvsr}(a), the accumulated mass grows rapidly, in particular for the interval between $r = 0$ and $r = 5$.  Beyond the central region, the accumulated mass increases less rapidly and becomes steady. The behavior of the accumulated mass is determined by the density and the pressure distribution within the star.  Initially, both the energy density and pressure in Figure~\ref{T=0,B=0RhoPvsr}, decrease rapidly then they drop to zero more gradually at larger distance.  The chemical potential also behaves similarly(Figure~\ref{T=0,B=0MMuvsr}(b)). It is clear that the matter in the star becomes extremely dense in the region near the core. Figure~\ref{T=0,B=0MasslimitMu} shows the mass curve of the degenerate star as a function of the central chemical potential and density.  From numerical analysis, the maximum mass is found to be $M_{max}=0.767302$ for the central chemical potential equal to $e^{1.033}$ or at the central energy density equal to $e^{-0.122306}$.  This maximal mass can be interpreted to be the mass limit above which gravitational collapse occurs.  A mass injection into an empty AdS space until the accumulated mass exceeds the mass limit would result in a gravitational collapse in the bulk.  The collapse corresponds to a thermalization process to finite temperature of the dual gauge matter.  Therefore, the mass limit corresponds to the minimum injected mass required by the dual gauge matter to start the thermalization process into the thermal equilibrium.  After deconfinement thermalization, the dual gauge matter is in thermal equilibrium at the Hawking temperature at this mass limit, i.e. $T_{gauge}=T_{H}$ with~\cite{hmr}
\begin{eqnarray}
T_{H} & = & \frac{1}{\pi \ell}\left(\frac{r_{+}}{\ell} \right)+\frac{1}{2\pi r_{+}}, \label{thm}
\end{eqnarray}
where the horizon radius $r_{+}=\ell\left( (\sqrt{1+4MC_{4}/\ell^{2}}-1)/2\right)^{1/2}$ for AdS$_{5}$.  Note that the mass dependence of the Hawking-Page temperature in the limit of large~($r_{+}\gg \ell$) and small~($r_{+}\ll \ell$) black hole in the AdS is
\begin{eqnarray}
T_{H} & \simeq & \frac{(M C_{4})^{1/4}}{\pi \ell^{3/2}}, \frac{1}{2\pi\sqrt{M C_{4}}} \end{eqnarray}
respectively.

It is interesting to note that for $r_{+}/\ell <\sqrt{1/2}$~(small black hole with negative specific heat after the gravitational collapse), the higher the mass limit, the smaller temperature the dual gauge matter would thermalize to.  This corresponds to $M <3\ell^{2}/4C_{4}=9\pi/32~$(for $\ell=1$, approximately 0.8836).  The mass limit of our AdS star for $T, B=0$ is roughly $0.767$ and therefore the black hole at the end of gravitational collapse for AdS star at this mass limit is a small black hole with small negative specific heat.

\subsection{Case II, zero temperature and finite magnetic field}

For this case, the magnetic field is turned on and the mass limit and other properties at zero temperature are studied by comparing to the results of Case I.  Since the changes from case I is small, we will present the results using the numerical differences between the two cases.  Starting from the pressure and energy density for nonzero magnetic field
    \begin{subequations}
    \begin{align}
        P =& \left(\frac{g_{s}\pi^{2}}{30c^{4}h^{4}}\right)\left(3\mu(r)^{5} - 10m^{2}c^{4}\mu(r)^{3} + 15m^{4}c^{8}\mu(r) - 8m^{5}c^{10} - 20m^{2}c^{4}\mu_{B}^{2}B^{2}(\mu - mc^{2})\right), \label{Pressure in Simulation at T=0} \\
        \rho c^{2} =& \left(\frac{g_{s}2\pi^{2}}{15c^{4}h^{4}}\right)\left(3\mu(r)^{5} - 5m^{2}c^{4}\mu(r)^{3} + 2m^{5}c^{10}\right) - mc^{2} \left(\frac{4m^{2}\pi^{2}\mu_{B}^{2}B^{2}}{3h^{4}}\right). \label{Energy density in Simulation at T=0}
    \end{align}
    \end{subequations}
 Observe that the pressure of the star has almost the same form as the pressure in Case I.  The correction term to the pressure from the magnetic field contains the factor $\mu - mc^{2}$.  The density appears to be smaller due to the contribution from the term $-mc^{2}\left(\frac{4m^{2}\pi^{2}\mu_{B}^{2}B^{2}}{3h^{4}}\right)$.

Since the pressure vanishes at $\mu = mc^{2}$ as in case I, the surface of the star is defined in the similar way, at $\mu(R)=mc^{2}$.  Interestingly at this radius, the density becomes negative
\begin{eqnarray}
\rho(R)& = & - \frac{4m^{3}c^{2}\pi^{2}\mu_{B}^{2}B^{2}}{3h^{4}},
\end{eqnarray}
due to the interaction energy between the fermion's magnetic moment and the external field.  Interestingly, there is a critical field strength where the density becomes zero,
\begin{eqnarray}
B_{c} & = & \frac{mc^{2}}{\mu_{B}}\sqrt{\frac{3u^{5}-5u^{3}+2}{5}},
\end{eqnarray}
where $u\equiv \mu/mc^{2}$ is a rescaled chemical potential.  For magnetic field stronger than this critical value, the energy density becomes negative and there is no star formation or black hole in the bulk.  Since there is no horizon in the bulk, the dual gauge matter is at zero temperature under extremely strong magnetic field.

Likewise, there is a critical field where the pressure becomes zero,
\begin{eqnarray}
B_{c}^{\prime} & = & \frac{mc^{2}}{\mu_{B}}(u-1)\sqrt{\frac{1}{20}(3u^{2}+9u+8)}.
\end{eqnarray}
For $u > 1$, $B_{c}^{\prime}$ is always smaller than $B_{c}$, therefore the pressure becomes negative before the density as the field is increased.  At $u = 1$, both $B_{c}$ and $B_{c}^{\prime}$ are zero.

    \begin{figure}
        \centering
        \subfigure[]{\includegraphics[width=0.45\textwidth]{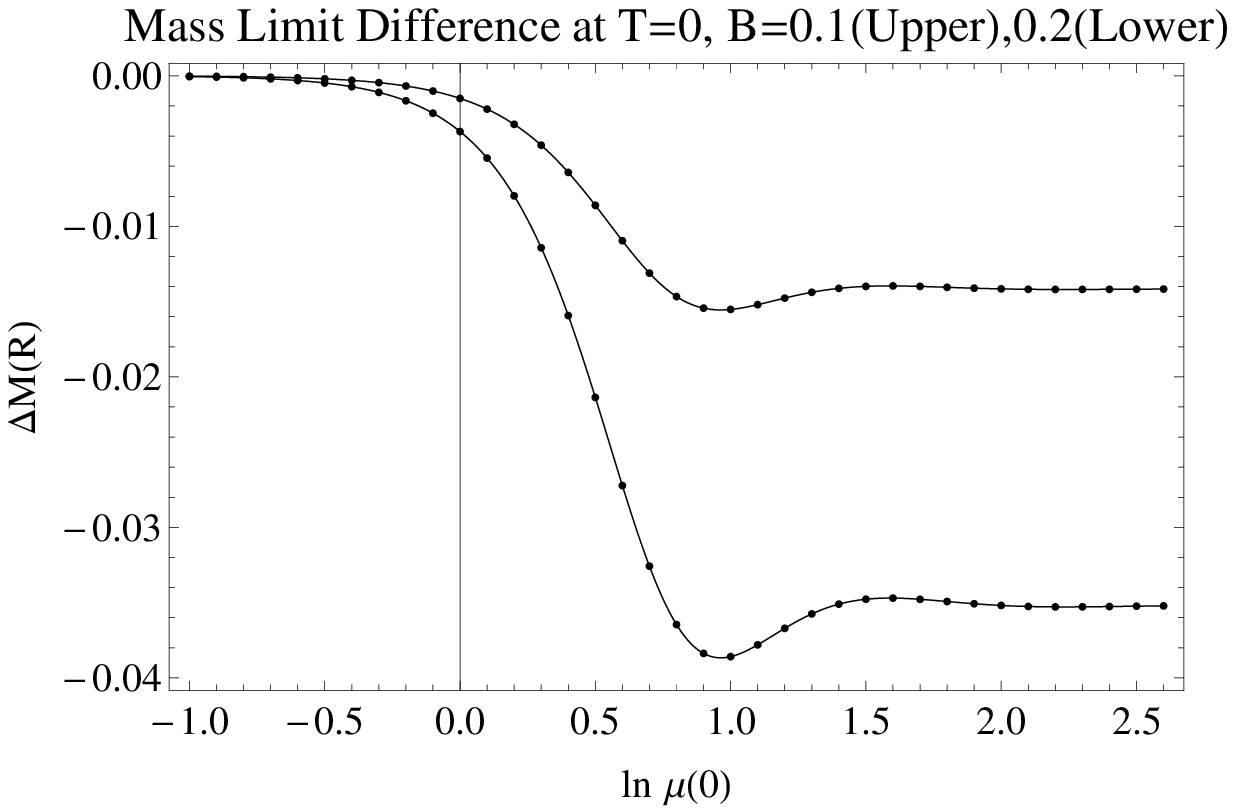}}\hfill
        \subfigure[]{\includegraphics[width=0.45\textwidth]{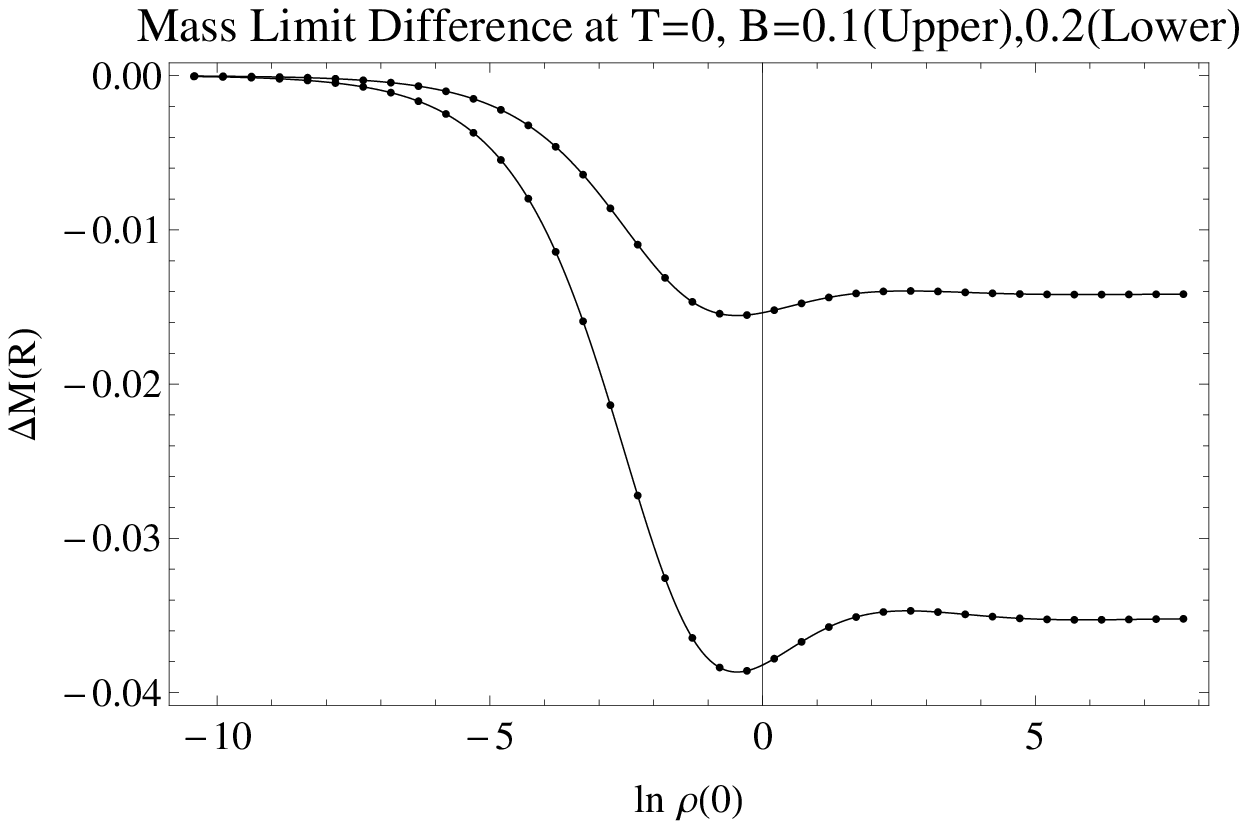}}
        \caption{The relation between mass and central chemical potential~(a) and central density~(b) of the degenerate star at $T = 0$, the mass difference between the nonzero magnetic field case and the $T,B=0$ case is presented.} \label{MixedatT=0MasslimitMu}
    \end{figure}
    \begin{figure}
        \centering
        \subfigure[]{\includegraphics[width=0.45\textwidth]{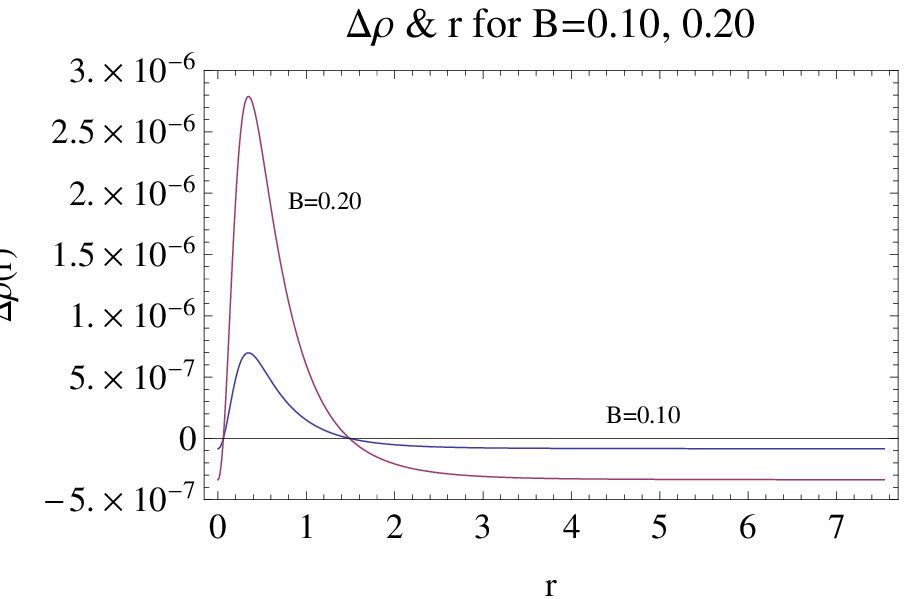}}\hfill
        \subfigure[]{\includegraphics[width=0.45\textwidth]{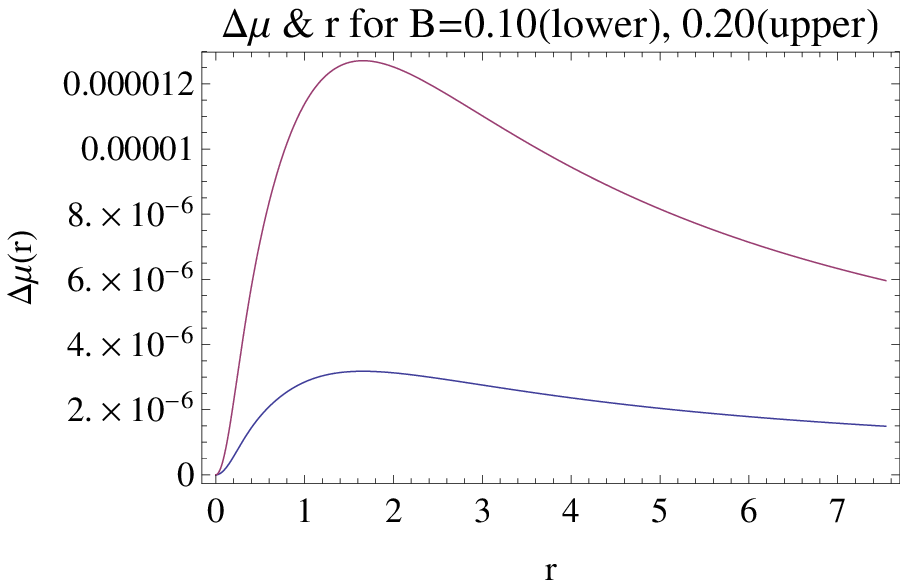}}\\
        \subfigure[]{\includegraphics[width=0.45\textwidth]{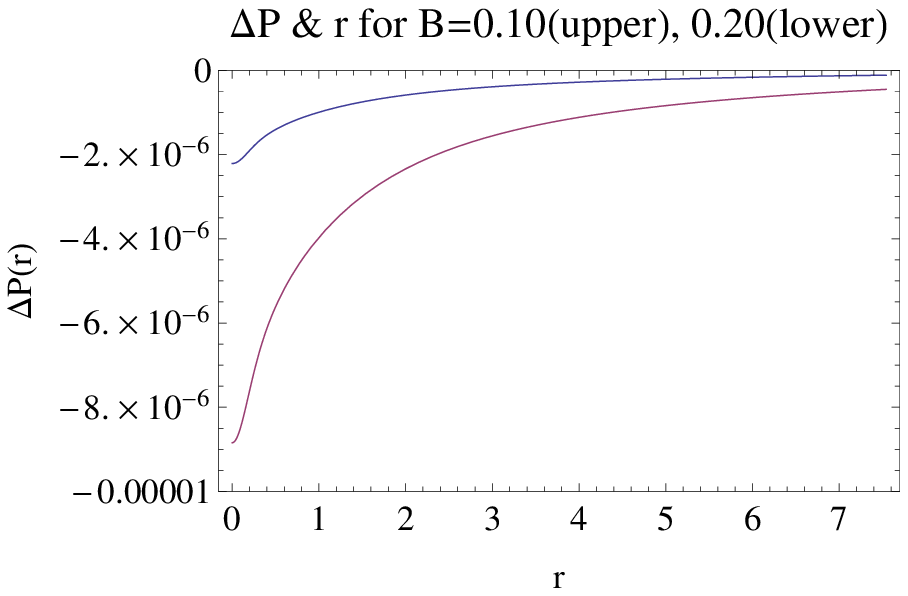}}
        \caption{The difference of the density~(a), the chemical potential~(b), the pressure~(c), between finite and zero magnetic field cases for $T=0$.} \label{OtherPropertiesatT=0Mvsr}
    \end{figure}

\newpage For numerical study, the magnetic field strength is chosen to be $0.10$ and $0.20$ for our consideration.  Fig.~\ref{MixedatT=0MasslimitMu} show that the mass limit, comparing to case I, decreases when the magnetic field increases.  The maximum mass for $B=0.2$ is appreciably smaller than the maximum mass at $B=0.1$.  Consider the equation of state in the energy density part (Eqn.~(\ref{Energy density in Simulation at T=0})). Since the coupled equations of motion between mass and chemical potential of the star (Eqn. (\ref{EoM of M}) and (\ref{EoM of mu})) involve the energy density, decreasing the energy density leads to the decrease of mass and the chemical potential of the star comparing to case I.  The increase of the chemical potential subsequently leads to the decrease in the pressure of the star.  Numerical analysis confirms these behaviour as are shown in Fig. \ref{OtherPropertiesatT=0Mvsr}.  Note that in the core region~($0<r\lesssim 1.4$), the density increases due to the increase of the chemical potential.  However, in the outer region of the star, the effect of the magnetic field becomes dominant resulting in the decrease of the density.  Accumulated mass eventually becomes smaller than the mass in case I.

The maximal mass or the mass limit of the AdS star when the magnetic field is turned on is smaller than the mass limit in case I.  Therefore the dual gauge matter under magnetic field thermalizes to larger temperature when the accumulated mass exceeds the mass limit even though it requires smaller injected mass in order to start the thermalization.  Gravitational collapse of an AdS star under strong magnetic field corresponds to thermalization of the magnetized gauge matter from zero to finite temperature.  Remarkably, the thermalized temperature~(at the mass limit) is {\it larger} than when the field is absent previously discussed in case I.  The magnetized gauge matter thermalizes more easily by requiring smaller injected mass, and also becomes hotter after the deconfinement thermalization.

\subsection{Case III, finite temperature and zero magnetic field}

For finite bulk temperature, the bulk fermions become thermal in the AdS space.  Since the kinetic energy of the particles increases, the pressure becomes larger and the star grows bigger.  Again, we study the small changes in the mass limit and other properties of the star by comparing the results to the zero temperature case.  The pressure and energy density, Eqn.~(\ref{Pressure in Simulation}),(\ref{Energy density in Simulation}) in this case reduce to
\begin{eqnarray}
P & = & \left(\frac{g_{s}\pi^{2}}{30c^{4}h^{4}}\right)\Bigg[3\mu(r)^{5} - 10m^{2}c^{4}\mu(r)^{3} + 15m^{4}c^{8}\mu(r) - 8m^{5}c^{10} - 10k_{B}^{2}T^{2}m^{2}c^{4}\pi^{2}\mu(r) \nonumber \\
&&  +~7k_{B}^{4}T^{4}\pi^{4}\mu(r) + 10k_{B}^{2}T^{2}\pi^{2}\mu(r)^{3} - 120k_{B}^{3}T^{3}m^{2}c^{4}Li_{3}\left(-e^{\frac{mc^{2}-\mu(r)}{k_{B}T}}\right)  \nonumber \\
&&  +~360k_{B}^{4}T^{4}mc^{2}Li_{4}\left(-e^{\frac{mc^{2}-\mu(r)}{k_{B}T}}\right) - 360k_{B}^{5}T^{5}Li_{5}\left(-e^{\frac{mc^{2}-\mu(r)}{k_{B}T}}\right) \Bigg], \label{Pressure in Simulation at B=0} \\
\rho c^{2} & = & \left(\frac{g_{s}2\pi^{2}}{15c^{4}h^{4}}\right)\Bigg[3\mu(r)^{5} - 5m^{2}c^{4}\mu(r)^{3} + 2m^{5}c^{10} - 5k_{B}^{2}T^{2}m^{2}c^{4}\pi^{2}\mu(r) + 7k_{B}^{4}T^{4}\pi^{4}\mu(r) \nonumber \\
&&  +~10k_{B}^{2}T^{2}\pi^{2}\mu(r)^{3} + 30k_{B}^{2}T^{2}m^{3}c^{6}Li_{2}\left(-e^{\frac{mc^{2}-\mu(r)}{k_{B}T}}\right) - 150k_{B}^{3}T^{3}m^{2}c^{4}Li_{3}\left(-e^{\frac{mc^{2}-\mu(r)}{k_{B}T}}\right)  \nonumber \\
&&  +~360k_{B}^{4}T^{4}mc^{2}Li_{4}\left(-e^{\frac{mc^{2}-\mu(r)}{k_{B}T}}\right) - 360k_{B}^{5}T^{5}Li_{5}\left(-e^{\frac{mc^{2}-\mu(r)}{k_{B}T}}\right)\Bigg]. \label{Energy density in Simulation at B=0}
\end{eqnarray}
It is interesting to investigate the large temperature limit, $k_{B}T \gg mc^{2}, \mu$.  In this limit, the polylogarithmic function becomes a zeta function $Li_{s}(-1)=-(1-2^{1-s})\zeta(s)$ and thus
\begin{eqnarray}
P& = & \left(\frac{g_{s}\pi^{2}}{30c^{4}h^{4}}\right) \frac{675}{2}\zeta(5)(k_{B}T)^{5}, \quad \quad\text{for large} ~~k_{B}T.
\end{eqnarray}
If we assume the star to be in a uniform temperature, this implies that the thermal fermions are not confined within a finite-size star when the temperature is sufficiently large, i.e. $k_{B}T \gg mc^{2}, \mu$.  The result is not surprising, any particles with sufficiently large kinetic energy will escape the gravitational influence of the star.

\begin{figure}[t]
        \begin{center}
          \subfigure[]{\includegraphics[width=0.45\textwidth]{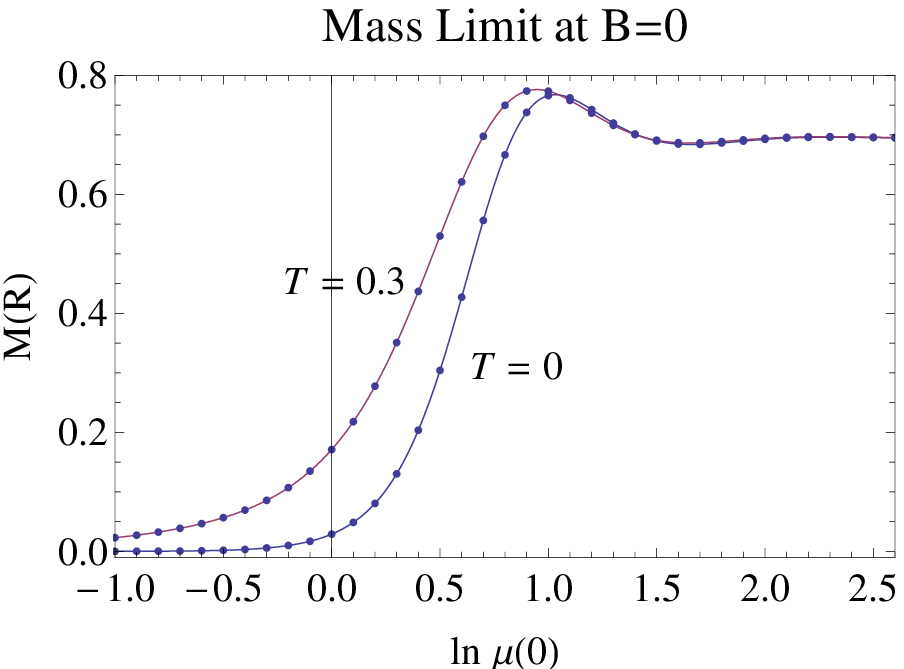}}\hfill
          \subfigure[]{\includegraphics[width=0.45\textwidth]{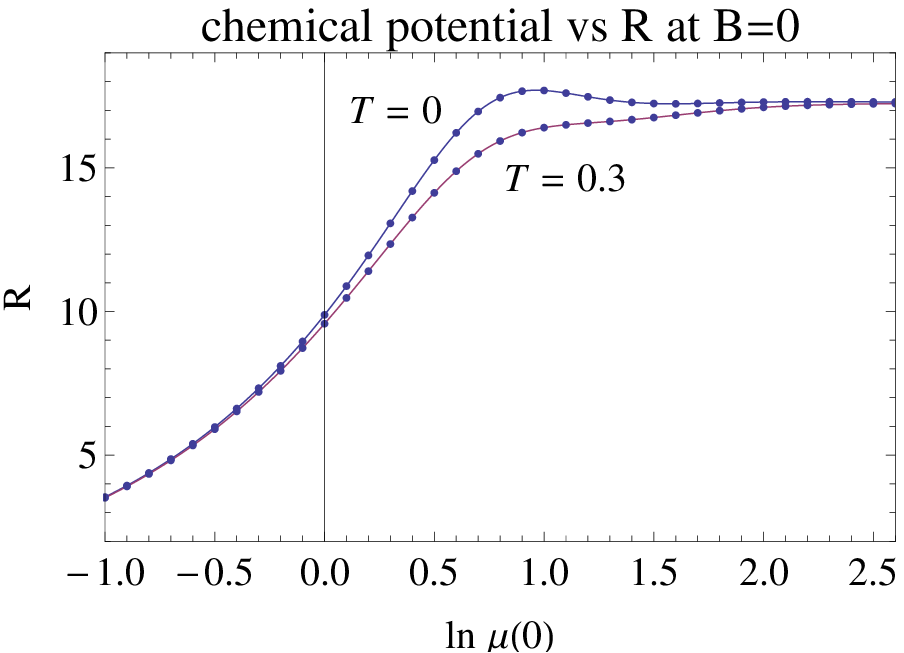}}
        \end{center}
        \caption{(a) The mass curves for $B=0$ as a function of the central chemical potential.  (b) The radius of the AdS star as a function of the central chemical potential for $B=0$. }
        \label{MixedatB=0MasslimitMu}
    \end{figure}

\begin{figure}[h]
        \begin{center}
          \subfigure[]{\includegraphics[width=0.45\textwidth]{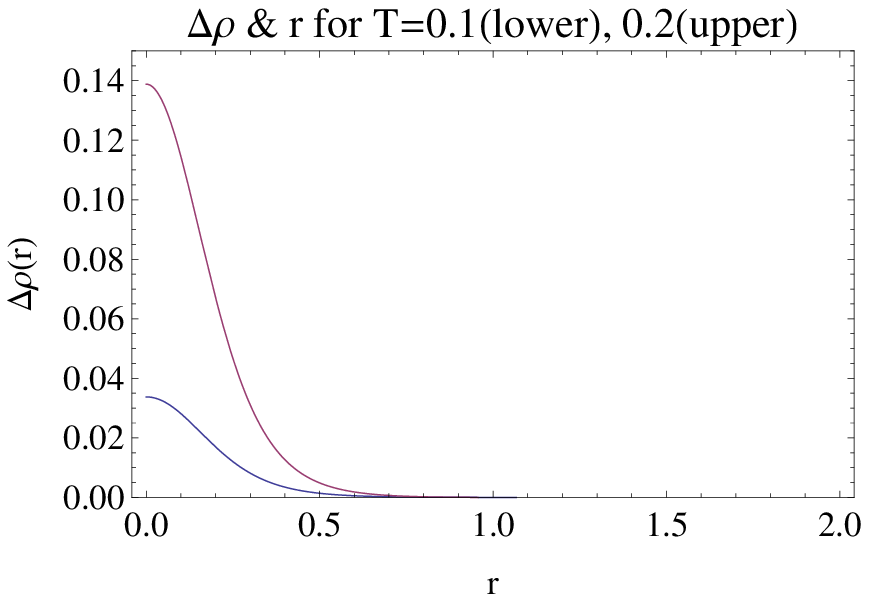}}\hfill
          \subfigure[]{\includegraphics[width=0.45\textwidth]{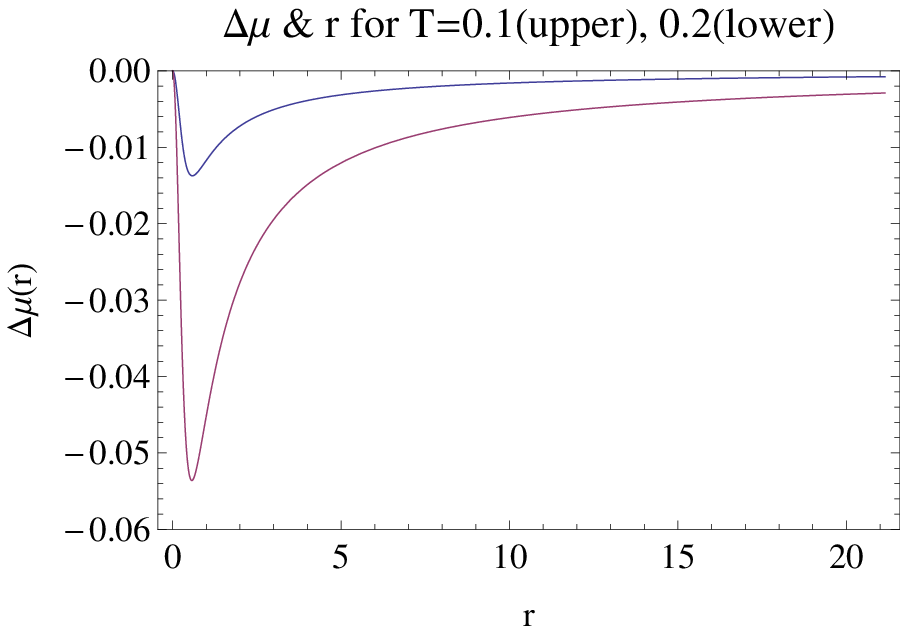}}\\
        \end{center}
        \caption{The difference of the density~(a), and the chemical potential~(b), between finite and zero temperature cases for $B=0$.} \label{OtherPropertiesatB=0Rhovsr}
    \end{figure}

We set temperature values in the simulation unit to be $0-0.3$.
Figure~\ref{MixedatB=0MasslimitMu} show that temperature increasing hardly affects the mass limit.  For this case, the surface of star is defined at $\mu(r=R)=mc^{2}$ since the density and pressure do not necessarily reduce to zero.  The maximum masses increase with the bulk temperature.  This is because the small increase in the temperature affects the Fermi-Dirac distribution very slightly.  Most particles are still in the same quantum states, mostly degenerate, and a very small part of the particles occupy higher energies than the Fermi energy and exert more pressure.  Increasing temperature thus results in a small increase of pressure and energy density. Consequently, when temperature increases, the maximum mass also grows.  For $T \gtrsim 0.1, 0.2$, the energy density and chemical potential reduce to zero at much larger radii as shown in Fig.~\ref{OtherPropertiesatB=0Rhovsr}.  For sufficiently large temperature, even though the chemical potential reduces to $mc^{2}$ at smaller radii, the pressure does not reduce to zero.  In other words, the thermal bulk fermions refuse to be confined within a finite-size star above a critical temperature.


To interpret the results in the dual gauge picture, caution has to be made regarding the bulk temperature.  During the thermalization process corresponding to the gravitational collapse in the gravity picture, the gauge matter is not in thermal equilibrium until a black hole is formed when the mass injection exceeds the mass limit.  A zero-bulk-temperature AdS star collapsing into a black hole becomes thermal at nonzero Hawking temperature due to the emergence of a horizon.  Therefore, the bulk temperature does not correspond to any sort of temperature of the gauge matter on the boundary world.  One of the effects of the bulk temperature of the fermions in the AdS star is the increase of mass limit.  Once a black hole is formed from gravitational collapse of the warm AdS star, the corresponding Hawking temperature is always smaller than the the zero bulk temperature case.  After thermalization process, the dual gauge matter will be in thermal equilibrium at {\it lower} temperature than the case of zero bulk temperature collapse.  However, the total injected energy is larger than the zero bulk temperature case.  The bulk temperature thus serves as a parameter which delays the onset of the thermalization process as well as reducing the temperature of the resulting thermal equilibrium.

Certainly, the dual gauge matter at exactly the same temperature can be alternatively achieved by injecting mass into a black hole in AdS space, increasing its mass and reducing the corresponding Hawking temperature~(however, if we keep increasing the black hole mass, it will finally become large black hole with positive specific heat and the temperature will start to increase with the mass).  This choice would correspond to in-equilibrium thermalization where the gauge matter is always kept at thermal equilibrium as temperature decreases.  The final thermal equilibrium at certain temperature can always be achieved by infinitely many different thermalization processes.

\subsection{Case IV, finite temperature and finite magnetic field}

We now consider effects from both the finite bulk temperature and nonzero magnetic field to the mass limit and other properties of the star. The equations of state have the full form according to Eqn.~(\ref{Pressure in Simulation}) and (\ref{Energy density in Simulation}).
    \begin{figure}
        \centering
        \includegraphics[width=0.60\textwidth]{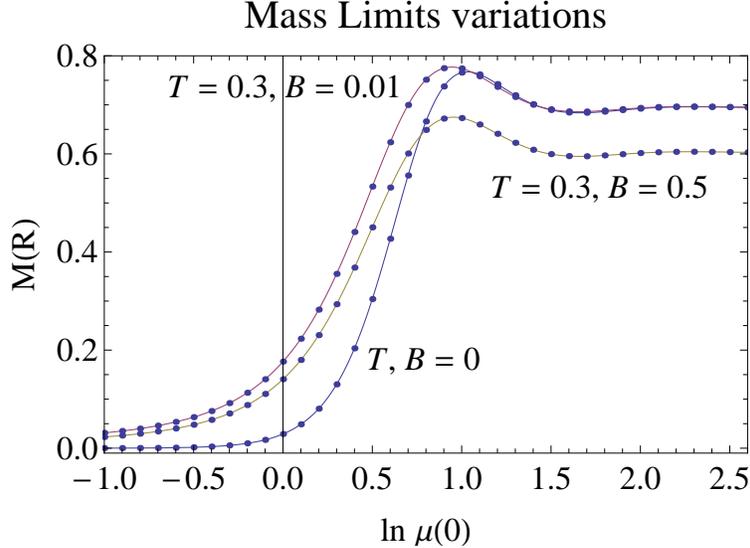}
        \caption{The mass limit curves for $T=0.3, B=0.01$ and $T=0.3, B=0.5$ in comparison to the mass limit curve at $T,B=0$.} \label{Mixed}
    \end{figure}
Again, it is interesting to consider the extreme limit of large temperature in the presence of the magnetic field.  For nonzero field, the pressure in this limit becomes
\begin{eqnarray}
P& = & \left(\frac{g_{s}\pi^{2}}{30c^{4}h^{4}}\right) \left(\frac{675}{2}\zeta(5)(k_{B}T)^{5} - 20 k_{B}Tm^{2}c^{4}\mu_{B}^{2}B^{2}\ln{2}\right), ~\text{for} ~~k_{B}T, \mu_{B}B \gg \mu, \label{largeBTP}
\end{eqnarray}
provided that the field $B$ is also comparably large.  From Eqn.~(\ref{largeBTP}), the star will have definite surface at finite radius when $B \propto T^{2}$.  Sufficiently hot star requires sufficiently strong field to confine its fermionic content.

We see the similar behaviour as in case II and III, temperature increase leads to the increase of the mass limit whereas the effect of the magnetic field is the opposite. In Fig. \ref{Mixed}, when we set the field $B = 0.01$, the temperature $T = 0.3$ has stronger effect on the profile of the star.  The mass limit becomes larger than the mass limit in the case of the zero temperature and magnetic field.  Similar to case III, when the temperature increases, the mass limit grows larger (the upper line in the Fig. \ref{Mixed}).  However, if the magnetic field is enhanced further to $B = 0.5$, the mass limit becomes smaller than the zero-field zero-temperature mass limit.  Namely, the influence of the magnetic field has overcome those of the temperature when it is sufficiently large.

Let us summarize implications for the thermalization of the dual gauge matter from the results in this mixed situation with $T,B > 0$.  Generically, turning on the bulk temperature results in a larger mass limit in the AdS space while finite magnetic field leads to a smaller mass limit.  If the injected mass exceeds the mass limit, gravitational collapse will occur and we end up with a black hole.  The injected mass at the mass limit is also the minimum mass required for the dual gauge matter to start the thermalization.  The Hawking temperature of the black hole can be identified with the temperature of the dual gauge matter at thermal equilibrium after the non-equilibrium thermalization process corresponding to the collapse, it is larger~(smaller) for finite field~(bulk temperature) than the collapse with $T,B = 0$.  The field and the bulk temperature compete with opposite effects.

For zero-field finite temperature collapse, the final black hole has higher mass and thus corresponds to small temperature of the gauge matter.  The final equilibrium at the same temperature can be achieved via in-equilibrium process by injecting mass into a black hole resulted from gravitational collapse of an AdS star with $T,B = 0$~(case I).  On the contrary, when the field is turned on, we need to extract mass from a magnetized black hole, reducing its mass and increasing its Hawking temperature in order to achieve the thermal equilibrium at the same temperature and magnetic field.

The black hole immersed in the constant magnetic field in 4 dimensions was originally investigated in Ref.~\cite{ernst}.  Extension to the magnetized black hole in AdS$_{5}$ spacetime is required to fully understand the holographic description of the strongly coupled magnetized gauge matter, one such solution~(magnetic brane) is discussed in Ref.~\cite{hk}.  It is found that the entropy density of the black brane in AdS$_{5}$ is proportional to $T$ for small $T$ and has a $T^{3}$ dependence for higher temperatures.  We will calculate the entropy density of the AdS star and compare to the case of magnetic brane in Section V.  However, as stated above, we have assumed the field is not sufficiently strong that it affects the spacetime of the background and our analyses are thus limited to the moderate magnetic field situation.

\section{Mass-radius relations}

\begin{figure}[h]
        \centering
        \subfigure[]{\includegraphics[width=0.45\textwidth]{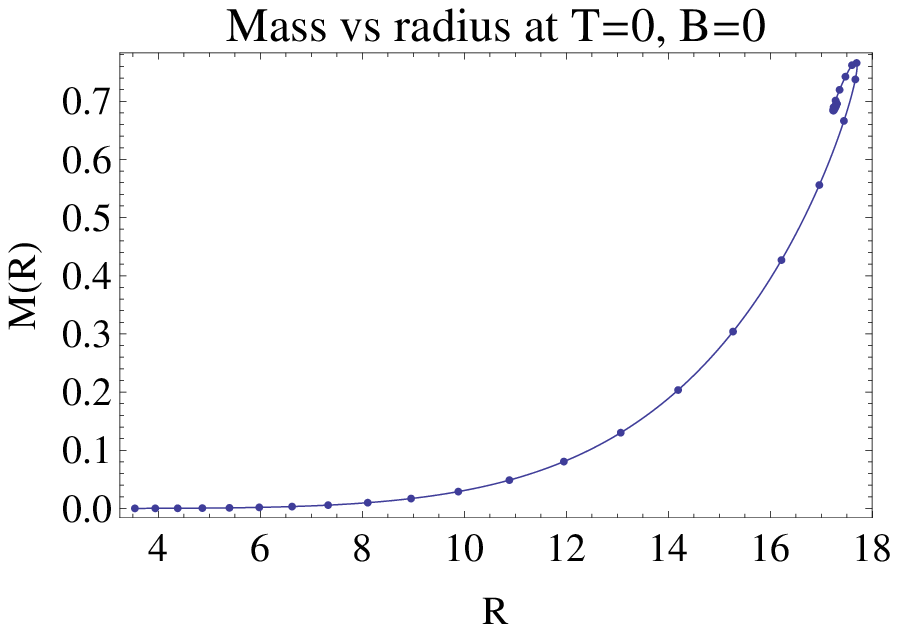}}\hfill
        \subfigure[]{\includegraphics[width=0.45\textwidth]{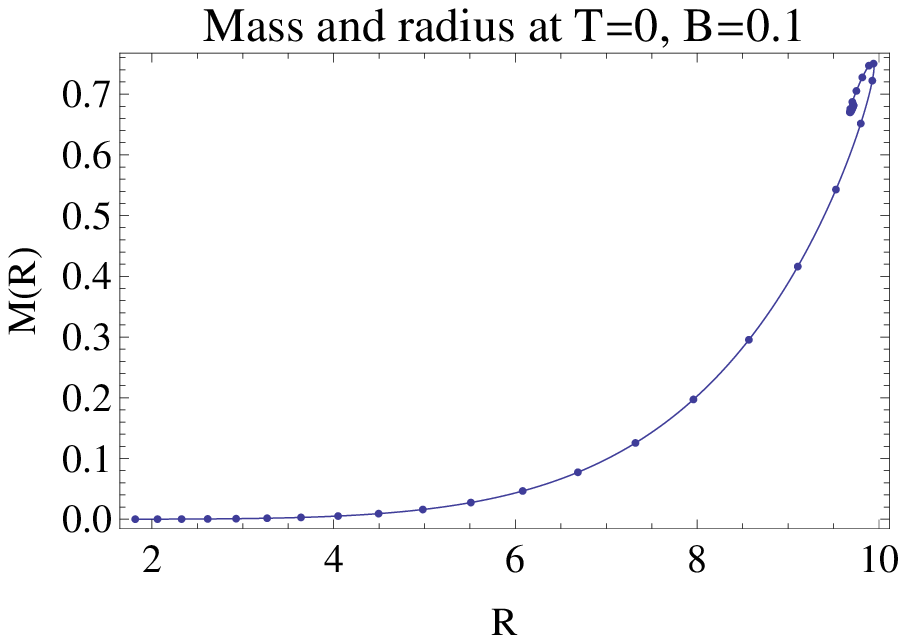}}\hfill
        \subfigure[]{\includegraphics[width=0.45\textwidth]{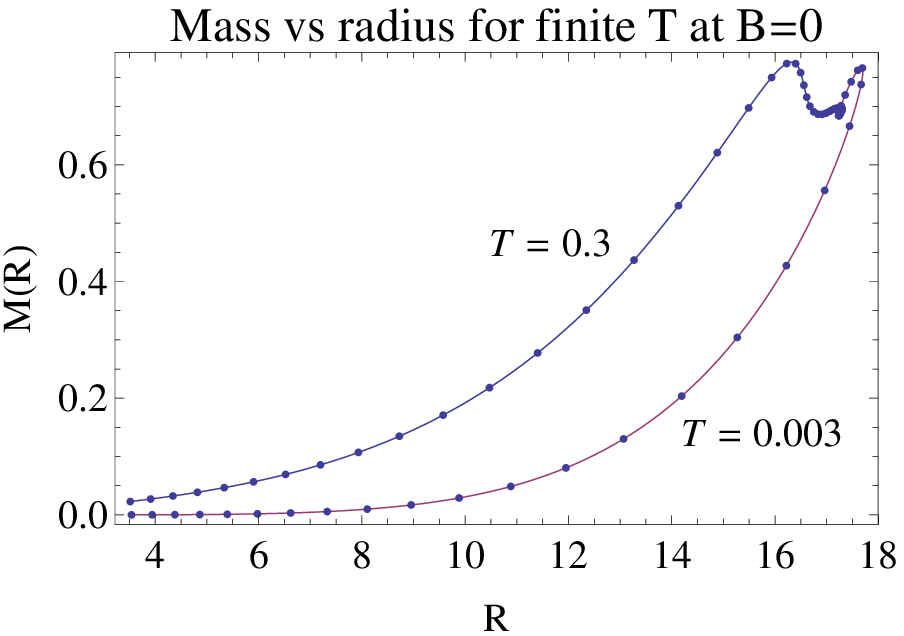}}\hfill
        \subfigure[]{\includegraphics[width=0.45\textwidth]{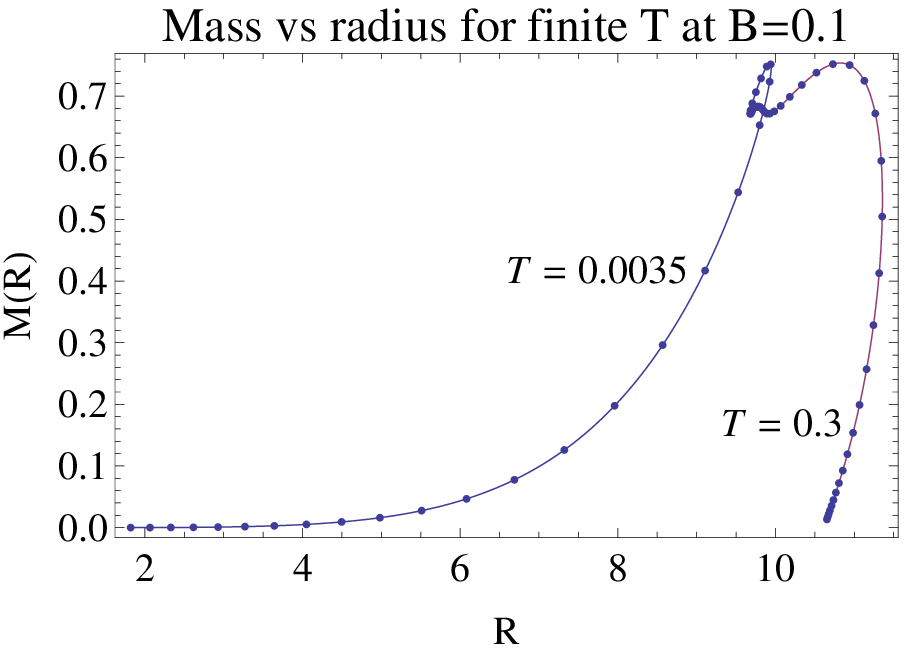}}
        \caption{Relationships between mass and radius of the fermionic star.} \label{MRrelation}
    \end{figure}

The mass sequences diagram of the AdS star for each case can be presented by the mass-radius plot of the star as shown in Fig.~\ref{MRrelation}.  Figure~\ref{MRrelation}(a) is the mass-radius sequence for case I with zero temperature and zero magnetic field.  The stars in this case have larger radius than case II~(zero temperature, finite field) in Fig.~\ref{MRrelation}(b) but smaller radius than case III~(zero field, finite temperature) in Fig.~\ref{MRrelation}(c).  Interesting competition between temperature and magnetic field can be seen in Fig.~\ref{MRrelation}(d), a sufficiently large field helps to confine the fermions within a finite-size star even for relatively higher temperatures comparing to case III.

For sufficiently high temperature, the mass-radius curve can change the way it spirals to the attractor fixed point at $\mu(0)\to \infty$.  For $B=0$ in Fig.~\ref{MRrelation}(c), the curve with $T=0.3$ ``oscillates down" to the fixed point from the small radii instead of the typical anticlockwise spiralling.  This is because at this temperature the radius of the star is an increasing function of $\mu(0)$ with no oscillation as we can see from Fig.~\ref{MixedatB=0MasslimitMu}(b).  For $B=0.1$ in Fig.~\ref{MRrelation}(d), the curve with $T=0.3$, ``oscillates down" to the fixed point from the large radii without spiralling. It should be remarked that for case I and III~(zero field), the mass at the attractor fixed point for $\mu(0)\to \infty$ is around 0.7.  For case II and IV at $B=0.1$, the fixed point mass for $\mu(0)\to \infty$ is around 0.68.  The radius of the AdS star at the fixed point decreases with the field but does not depend very sensitively on the temperature.

\section{The adiabatic index, sound speed, entropy density and total entropy of the AdS star}

Many interesting physical properties of the fermions squeezed within the AdS star by its own gravity can be illustrated by certain thermodynamic and transport quantities.  In this section, we consider two transport coefficients, the adiabatic index and sound speed of the AdS fermionic matter for each limiting case.  The entropy density and total entropy of the AdS star are discussed subsequently.

Generically the adiabatic index, $\Gamma$, and the sound speed, $c_{s}$, of a medium are defined as
\begin{eqnarray}
\Gamma & = & \frac{\rho}{P}\frac{\partial P}{\partial \rho} = \frac{\rho}{P}c_{s}^{2}, \label{Gamma}  \\
       & = & \frac{\rho}{P}\frac{\partial_{\mu} P}{\partial_{\mu} \rho} \nonumber
\end{eqnarray}
which can be calculated through the dependence on the chemical potential $\mu$ of both $P$ and $\rho$.  The general expressions for both quantities are very lengthy but they are simplified for the zero-temperature limit.

$\underline{\text{For $T=0$, finite $B$}}$,
\begin{eqnarray}
\Gamma & = & \frac{3(u^{2}-1)^{2}-4v^{2}}{3u^{2}(u-1)^{2}(u+1)}\left(\frac{3u^{5}-5u^{3}+2-5v^{2}}{(3u^{2}+9u+8)(u-1)^{2}-20v^{2}}\right),  \\
c_{s}  & = & \frac{1}{2}\sqrt{\frac{(u^{2}-1)^{2}-\frac{4}{3}v^{2}}{u^{2}(u^{2}-1)}}, \label{cs}
\end{eqnarray}
where $u\equiv \mu/mc^{2}$ is the rescaled chemical potential and $v\equiv \mu_{B}B/mc^{2}$ is the rescaled magnetic energy of the fermions.

$\underline{\text{For $T,B=0$}}$,
\begin{eqnarray}
\Gamma & = & \left(\frac{1+u}{u^{2}}\right)\frac{3u^{3}+6u^{2}+4u+2}{3u^{2}+9u+8},  \\
c_{s}  & = & \frac{1}{2}\sqrt{1-\frac{1}{u^{2}}}.  \label{csB0}
\end{eqnarray}
A number of remarks are in order for the zero-temperature limit.  From Eqn.~(\ref{cs}) and (\ref{csB0}), the sound speed for the nonzero field case~($v^{2}>0$) is shown to be larger than the case with $B=0$.  For $B=0$ since $\mu \geq mc^{2}~(u\geq 1)$, the sound speed is always real and the upper limit of $c_{s}$ is always smaller than $1/2$ or half the speed of light.  For nonzero field, reality condition of $c_{s}$ leads to the constraint $v\leq \sqrt{3}(u^{2}-1)/2$.  Namely, for a given $u$, the upper limit on the magnetic field for ordinarily-compressible fermionic matter is
\begin{eqnarray}
B_{0} & = & \frac{\mu}{\mu_{B}}\frac{\sqrt{3}}{2}(u^{2}-1).
\end{eqnarray}
On the other hand, the upper limit from the light speed $c_{s}\leq 1$ is satisfied trivially for any value of $B$.

\begin{figure}[h]
        \centering
        \subfigure[]{\includegraphics[width=0.45\textwidth]{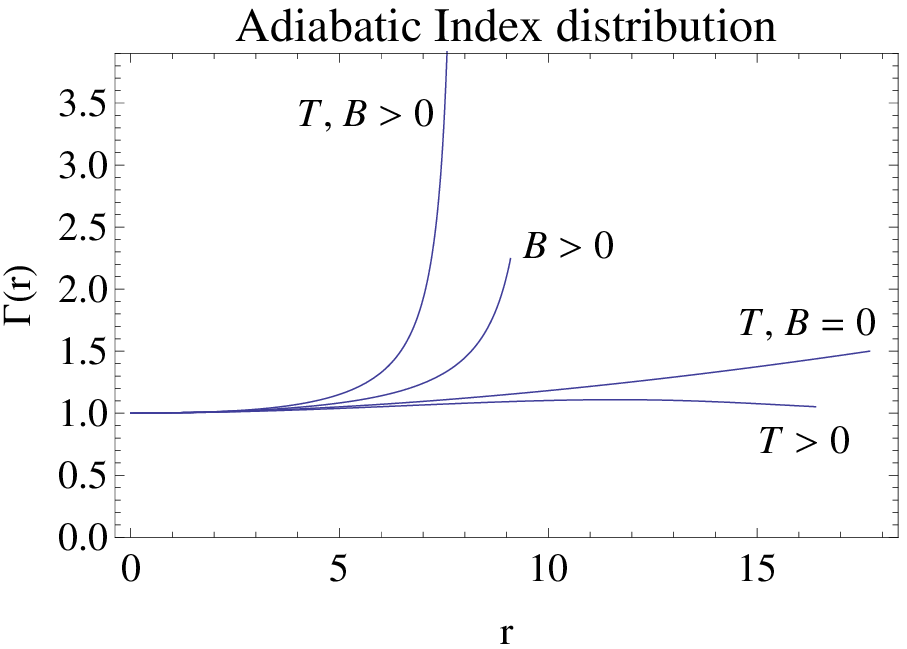}}\hfill
        \subfigure[]{\includegraphics[width=0.45\textwidth]{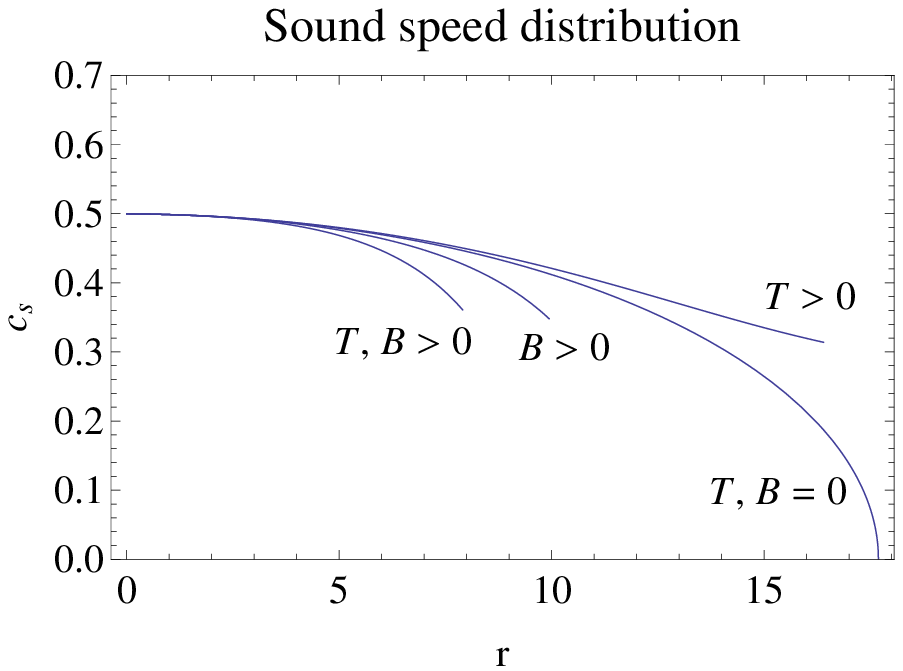}}
        \caption{The adiabatic index and sound speed of the fermionic matter in the AdS star.  The $T,B > 0$ label represents $T = 0.3, B = 0.2$ curve, the $T > 0$ and $B > 0$ label represents $T = 0.3, B = 0$ and $T = 0, B = 0.1$ curve respectively.} \label{gcs}
    \end{figure}

Numerical results for each case are presented in Fig.~\ref{gcs}.  The $B > 0$ and $T > 0$ label represents the curve with $T = 0, B = 0.1$ and $B = 0, T = 0.3$ respectively.  The $T,B > 0$ label represents the curve with $T = 0.3, B = 0.2$.

When a thermodynamical system is injected with energy until it reaches a thermal equilibrium, the total energy density, pressure and number density are related to the entropy density by the relation $sT=P+\rho-\mu n$ where the entropy density in our case can be computed via
\begin{eqnarray}
s & = & \frac{\partial P}{\partial T}\Big{\vert}_{\mu},
\end{eqnarray}
from the Gibbs-Duhem relation.  Using Eqn.~(\ref{Pressure in Simulation}), the entropy density of the fermionic content of the AdS star at finite temperature can be calculated to be
\begin{eqnarray}
s & = & \frac{4 \pi ^2}{15 c^4 h^4} \Bigg[ \frac{5 B^2 c^4 \mu_{B}^2 m^2 \left(\mu (r)-m c^{2}\right) e^{\frac{\mu (r)}{k_{B} T}}}{T \left(e^{\frac{c^2 m}{k_{B} T}}+e^{\frac{\mu (r)}{k_{B} T}}\right)}
-5 B^2 c^4 k_{B} \mu_{B}^2 m^2 \ln \left(e^{\frac{\mu (r)-c^2 m}{k_{B} T}}+1\right)  \nonumber \\
&&  -450 k_{B}^5 T^4 Li_{5}\left(-e^{\frac{c^2 m-\mu (r)}{k_{B} T}}\right)+360 c^2 k_{B}^4 m T^3 Li_{4}\left(-e^{\frac{c^2 m-\mu (r)}{k_{B} T}}\right)  \nonumber \\
&&  +90 k_{B}^4 T^3 \left(c^2 m-\mu (r)\right) Li_{4}\left(-e^{\frac{c^2 m-\mu (r)}{k_{B} T}}\right)
-90 c^2 k_{B}^3 m T^2 \left(c^2 m-\mu (r)\right) Li_{3}\left(-e^{\frac{c^2 m-\mu (r)}{k_{B} T}}\right) \nonumber \\
&&  -90 c^4 k_{B}^3 m^2 T^2 Li_{3}\left(-e^{\frac{c^2 m-\mu (r)}{k_{B} T}}\right)
 +30 c^4 k_{B}^2 m^2 T \left(c^2 m-\mu (r)\right) Li_{2}\left(-e^{\frac{c^2 m-\mu (r)}{k_{B} T}}\right)
 \nonumber \\
&&  +7 \pi^{4} k_{B}^4 T^3 \mu (r)+5 \pi ^2 k_{B}^2 T \mu (r)\left( \mu (r)^{2}-(m c^2)^{2} \right) \Bigg]. \label{sD}
\end{eqnarray}
The entropy density of the fermion gas approaches zero as $T\to 0$, a typical behaviour from a quantum ensemble satisfying the third law of thermodynamics.  In the low temperature limit, the last two terms of Eqn.~(\ref{sD}) remain dominant and thus
\begin{eqnarray}
s & \simeq & \frac{4\pi^{4}k_{B}\mu (r)}{15(hc)^{4}} \left[5 k_{B}T(\mu^{2}-m^{2}c^{4}) + 7\pi^{2}(k_{B}T)^{3} \right].
\end{eqnarray}
It is interesting to compare the $T$-dependence of our entropy density with the magnetized black hole studied in Ref.~\cite{hk} where $s\sim T$ for small temperatures and $s\sim T^{3}$ for larger temperatures.  In our case of the fermions in the AdS star, the origin of the temperature dependence is the typical behaviour of free relativistic fermi gas persisting in any dimensions.  For the magnetized AdS black hole, the entropy is determined from the central charge of the AdS$_{3}$ subspace of AdS$_{3}\times T^{2}$ interpolating with the AdS$_{5}$.  However, it must be aware that the bulk temperature of the AdS star and the Hawking temperature of the black hole are two distinct kinds of temperature.  Only the latter corresponds to the temperature of dual gauge matter at a thermal equilibrium.

The entropy density of the magnetized fermion gas given by Eqn.~(\ref{sD}) also depends on the magnetic field $s\sim B^{2}$.  The dependence nevertheless vanishes in the $T\to 0$ limit.  However, this formula is the result of the Euler-Maclaurin formula which is a good approximation for $k_{B}T\gg \mu_{B}B$, i.e. sufficiently high temperature.  For smaller temperatures, starting with Eqn.~(\ref{parf}), the zeroth mode becomes dominant and the field-dependence becomes $s\sim \partial_{T}\ln Z \sim B$.  This is also similar to the behaviour of the magnetized black brane~\cite{hk}.

\begin{figure}[h]
        \centering
        \includegraphics[width=0.6\textwidth]{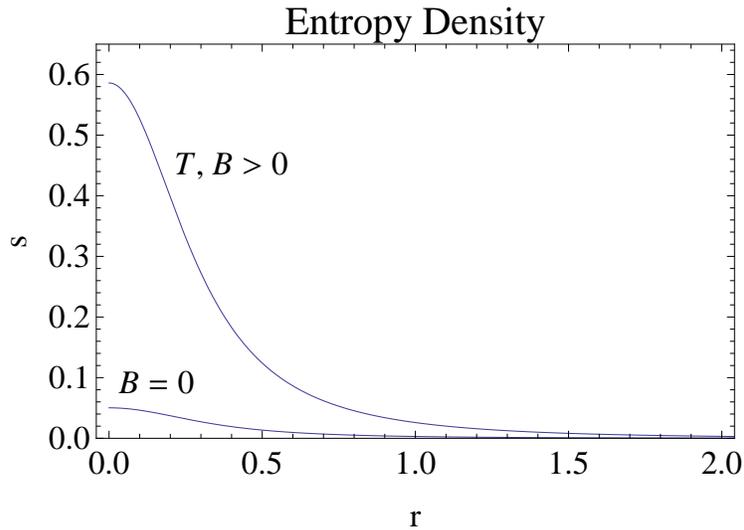}
        \caption{The entropy density of the fermionic matter in the AdS star.  The $T,B > 0$ label represents $T = 0.3, B = 0.2$ curve and the
         $B = 0$ label represents $T = 0.03, B = 0$ curve respectively.} \label{figsD}
    \end{figure}

The entropy density from two numerical solutions are shown in Fig.~\ref{figsD}.  Since $s$ is an increasing function of $T$, the star at relatively small temperatures will have smaller entropy density.  For small temperatures or small entropy density, we can approximate $sT \ll P, \rho, \mu n$ leading to $P+\rho \simeq \mu n$.

    \begin{figure}[h]
        \centering
        \includegraphics[width=0.45\textwidth]{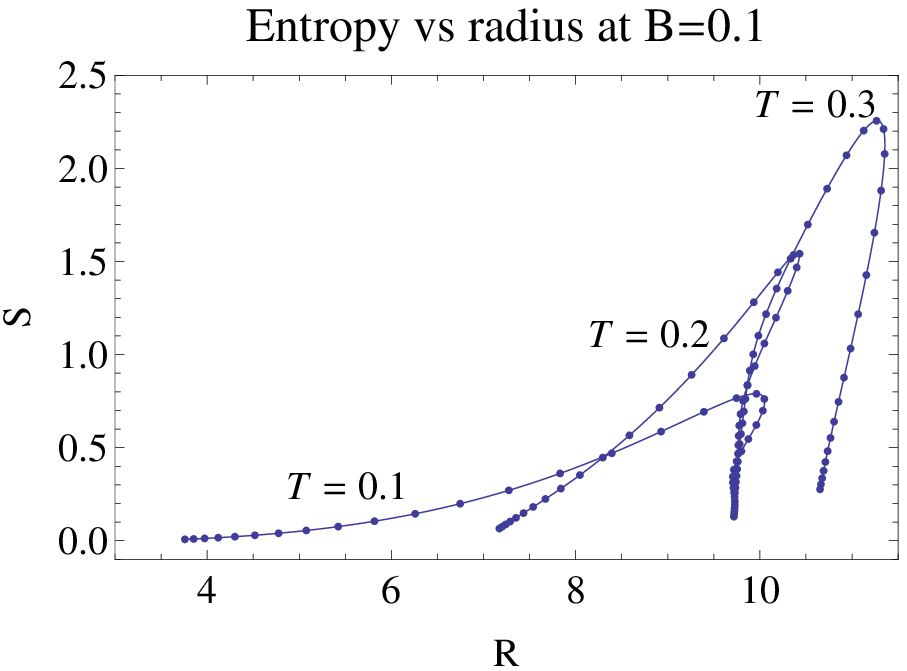}\hfill
        \includegraphics[width=0.45\textwidth]{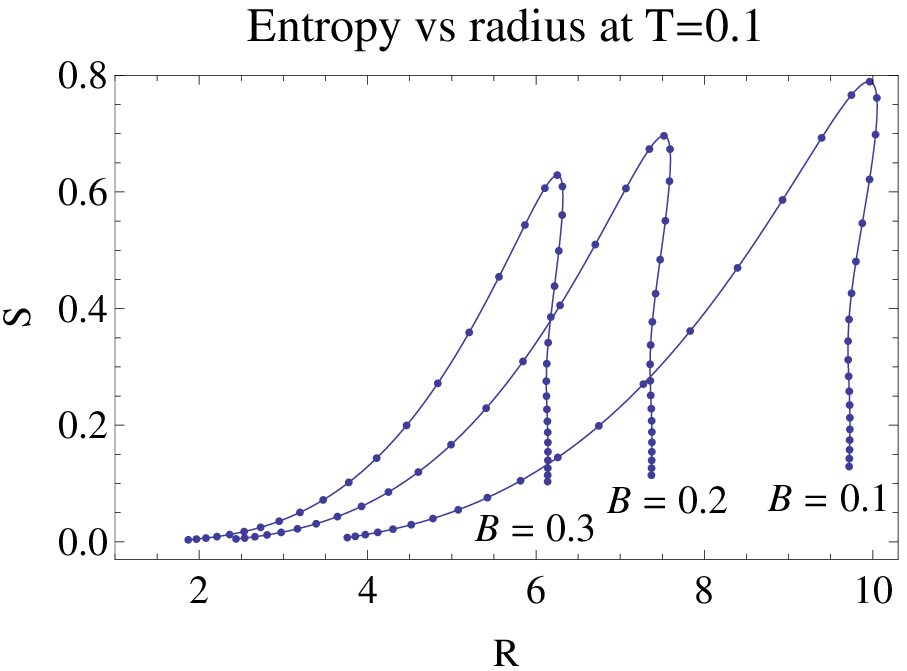}
        \caption{The total entropy as a function of radius of the AdS star for $B=0.1; T=0.1,0.2,0.3$ and $B=0.1,0.2,0.3; T=0.1$.} \label{figS}
    \end{figure}
    \begin{figure}[h]
        \centering
        \includegraphics[width=0.45\textwidth]{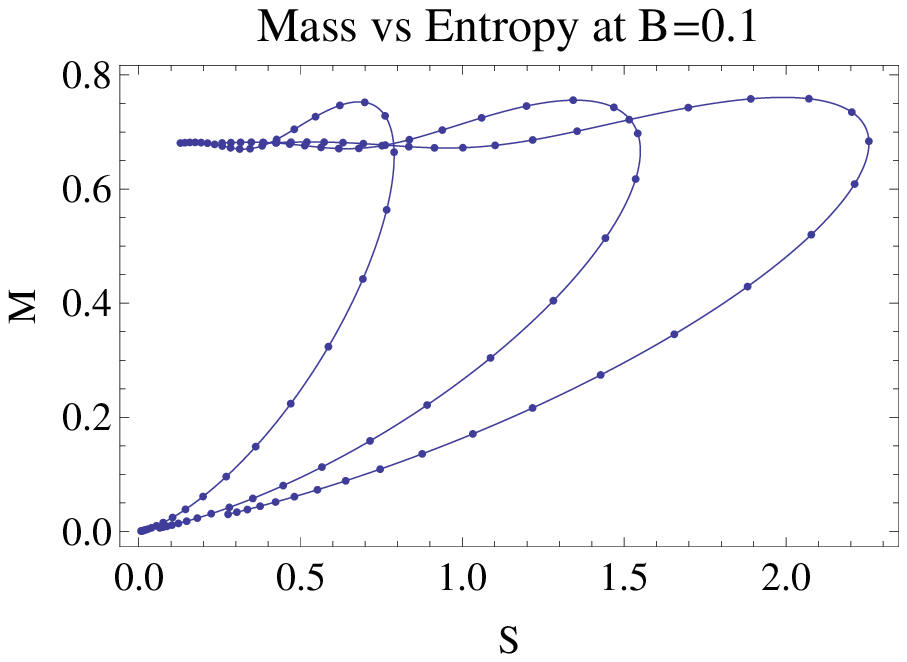}\hfill
        \includegraphics[width=0.45\textwidth]{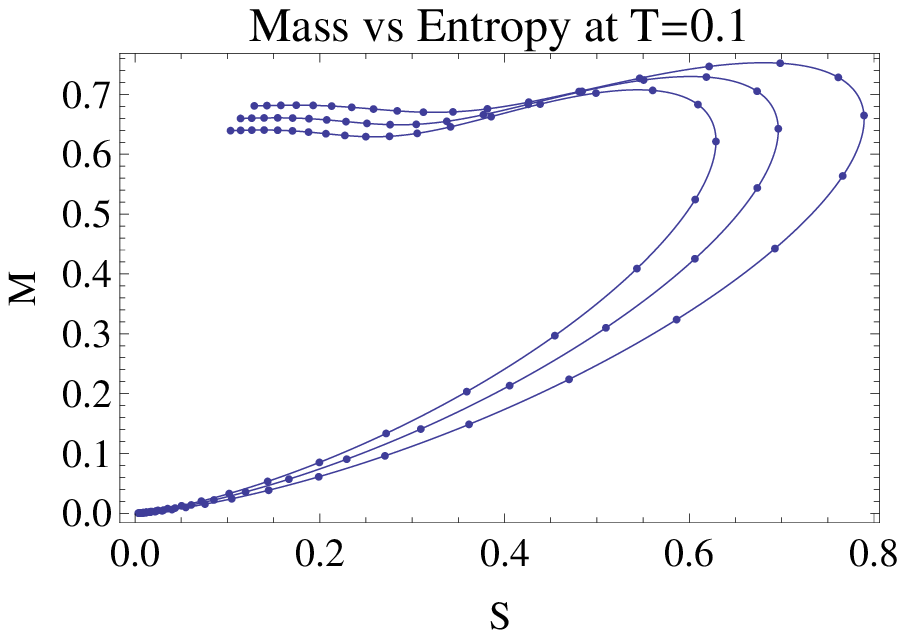}
        \caption{The total entropy as a function of mass of the AdS star for $B=0.1; T=0.1,0.2,0.3$~(from left to right) and $B=0.1,0.2,0.3; T=0.1$~(from right to left).} \label{figS1}
    \end{figure}

Next we calculate the total entropy of the AdS star which should be equivalent to the entropy of the dual gauge matter before the thermalization.  The total entropy of the star should be the lower bound of the total entropy of the black hole at the end of gravitational collapse when the mass of the AdS star exceeds the mass limit.  This black hole entropy in turn corresponds to the total entropy of the dual gauge matter at the end of thermalization.  In $d$ dimensions, the total entropy is given by
\begin{eqnarray}
S & = & \int_{0}^{R}~s(r)~\frac{2V_{d-2}}{d-2}~r^{d-2}~dr, \label{stot}
\end{eqnarray}
where the volume factor $2V_{d-2}r^{d-2}/(d-2)$ becomes $4\pi^{2}r^{3}/3$ for $d=5$.

Figure~\ref{figS} shows the total entropy of the AdS star for $B = 0.1$ at temperature $T = 0.1, 0.2, 0.3$ and for $T = 0.1$ under field $B = 0.1, 0.2, 0.3$.  The total entropy is an increasing function of the temperature and a decreasing function of the magnetic field.  From small radii, the total entropy is an increasing function of the star radius.  This is a similar behaviour to the accumulated mass which is also a global quantity.  Remarkably, the total entropy converges to zero in the attractor fixed point $\mu(0)\to \infty$ limit.  As the central density grows, the content of the AdS star concentrates more in the central region resulting in the decrease of total entropy towards zero~(the volume weighing factor $r^{3}$ enhances contribution in the outer region in contrast to the core).

The black hole at the end of gravitational collapse should possess at least the same amount of total entropy as the initial AdS star above the mass limit.  The second law of thermodynamic demands that the entropy of the AdS star above the mass limit is always less than the black hole entropy after the collapse~\cite{bek}.  The entropy increase could continue until it reaches the maximum when a black hole is formed~\cite{eg}.  Unfortunately, the time evolution of the entropy during the gravitational collapse is not completely known.  Partially because the thermal entropy is ill-defined during off-equilibrium processes and partially due to the geometric nature of black hole entropy at the end of the collapse.  There are other kinds of entropy that can be assigned to the AdS star and the black hole.  The entanglement entropy quantifies how much we do not know about the region behind the horizon and it is consistent with the geometric nature of the Bekenstein-Hawking entropy of the black hole.  Entanglement entropy is found to increase in a different manner from the Kolmogorov-Sinai entropy~\cite{bbb,bbbc} during the collapse.  However, all kinds of entropy are found to increase approximately linearly during the initial state of the gravitational collapse and saturate to a constant value at the end.

Each maximum of the $M$-$S$ curve in Fig.~\ref{figS1} corresponds holographically to the entropy of the dual gauge matter at the beginning of the thermalization process.  It is also proportional to entropy of the black hole after gravitational collapse assuming the linear progression to the saturated Bekenstein-Hawking entropy mentioned above.  Arguably, the increase of entropy of the dual gauge matter from the injected mass state to the thermal equilibrium should also be the linear progression following by saturation as well.  Note that the entropy is not maximal at the maximal mass nor the maximal radius as we can see from Fig.~\ref{figS}, \ref{figS1}.

\section{Dependence of mass limit on the AdS radius}

\begin{figure}
        \centering
        \includegraphics[width=0.5\textwidth]{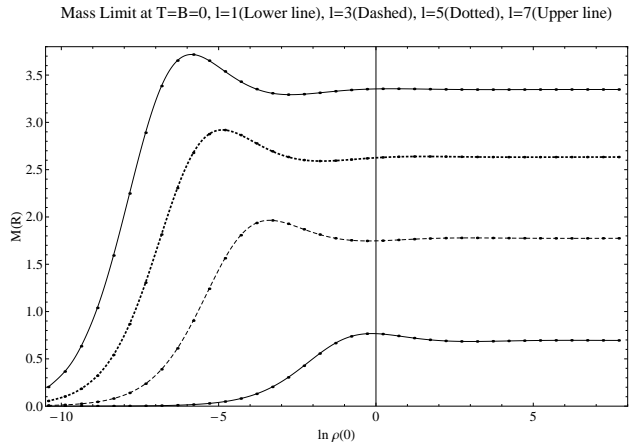}
        \caption{The relation between mass and central energy density (in logarithmic scale) of the degenerate star for varying AdS radius $l=1-7$.} \label{Mixed3}
    \end{figure}

We vary the curvature radius of the AdS space, $l$, and study the changes in the profile of the star in this section.  For simplicity, we will set the temperature and the external magnetic field to be zero.  We let the curvature radius to be $1, 3, 5$ and $7$, and observe considerable changes in the mass limit of the star as are shown in Fig. \ref{Mixed3}.
The mass limit of the degenerate star increases evidently when we raise the curvature radius of the AdS space.  Moreover, the peak of the mass limit curve shifts to the lower central density side.  Note that increasing $l$ corresponds to decreasing the bulk cosmological constant $\Lambda$.  For $l = 3$, the maximum mass is $1.96473 (r = 27.4029)$ for the central chemical potential $\mu(0) = e^{0.3825}$ or the central energy density $\rho(0) = e^{-3.38048}$. For $l = 5$, the maximum mass is $2.92023 (r = 33.5921)$ for the central chemical potential $\mu(0) = e^{0.083}$ or the central energy density $\rho(0) = e^{-4.88441}$. For $l = 7$, the maximum mass is $3.71782 (r = 38.4035)$ for the central chemical potential $\mu(0) = e^{-0.1115}$ or the central energy density $\rho(0) = e^{-5.86373}$.

\section{Conclusions and Discussions}

In this work we have found that both temperature and external magnetic field affect the mass limit and other physical properties of the fermionic AdS star. The increase of bulk temperature enables the pressure and energy density of the star to increase.  Consequently, the mass limit becomes slightly greater due to the larger pressure.  This is the typical behavior of the Fermi gas at finite temperature.  Too large temperature results in the the fermions refusing to be confined within a finite-size star, they will leak to the space inevitably.

In the presence of external magnetic field, the mass limit decreases when the magnetic field increases. As we can see from Eqn.~(\ref{Pressure in Simulation}) and (\ref{Energy density in Simulation}), an increase in the magnetic field results in a smaller energy and pressure density as well as a smaller chemical potential.  The mass limit becomes smaller naturally.  There is an interesting competition between the temperature and the magnetic field to the density profile and mass limit of the star.  Extremely strong magnetic field tends to make the bulk fermions stay in the Landau states with lower energies whilst the temperature causes the particles to flee the star.

The radius of curvature of the AdS space also affects the mass limit evidently.  When the radius of curvature increases, the mass limit increases substantially as are shown in Fig.~\ref{Mixed3}. Interestingly, the peak of the mass limit curve shifts to the lower central density side.

Gravitational collapse in the AdS space has holographic dual in terms of the non-equilibrium thermalization of the gauge matter on the boundary.  Even though the Hawking temperature of the black hole at the end of the gravitational collapse can be matched with the temperature at thermal equilibrium of the gauge matter at the end of thermalization, the bulk temperature of the AdS star does not seem to have such a straightforward relationship with the dual gauge matter.  The Hawking temperature of the resulting black hole is not directly related to the temperature of the fermionic star before the collapse but inversely proportional to the mass of the star.  Therefore it is the mass limit studied in our work which corresponds to the temperature of the gauge matter at the thermal equilibrium after thermalization, i.e. $T_{\text{gauge}}\sim 1/\sqrt{3/32\pi M_{\text{limit}}}$~(the black hole formed at our mass limit is small AdS black hole with negative specific heat, the precise relationship is given in Eqn.~(\ref{thm})), for a given mass injection $M_{\text{limit}}$ in the dual gauge picture.  The mass limit also plays the role of the minimum injected mass required for the dual gauge matter to undergo the thermalization into the thermal equilibrium.  Larger mass limit means that it requires more injected energy to thermalize, and once it thermalizes, the gauge matter will be at lower Hawking temperature.

It should be remarked that the AdS black holes formed at the end of the gravitational collapse of the AdS stars at our mass limits are small black holes with negative specific heat. They are previously thought to be less thermodynamically preferred than the AdS vacuum in the context of the AdS/CFT correspondence and only the large AdS black hole with positive specific heat was considered relevant for the dual of the thermal gauge matter.  However, inevitable collapse at the mass limits corresponding to small AdS black holes suggest that there might exist the phase of thermal gauge matter with negative specific heat dual to these black holes at the end of the gravitational collapse.  Injecting more mass would make these AdS black holes and their gauge duals eventually become thermodynamically stable with positive specific heat.

Entropy density of the AdS star under uniform magnetic field is found to show interesting behaviour; $s\sim T$ for small and $s\sim T^{3}$ for higher {\it bulk} temperatures.  Such $T$-dependence is typical for free fermion gas~(modulo the magnetic field existence) and it is amusingly similar to the $T$-dependence of the magnetic black brane entropy in the AdS~\cite{hk} even though the latter is the Hawking-Page temperature of the brane, not the bulk temperature of the material making up the brane itself.  Nevertheless, the correspondence between the bulk and boundary exists throughout the gravitational collapse as long as the background is the AdS.  The holographic duality suggests that the Hilbert spaces of both the gravity and gauge theory as well as their partition functions are equivalent.  A global probe for the number of degrees of freedom on both sides of the duality is the entropy.  The total entropy of the AdS star above the mass limit, which indicates the lower bound of the black hole entropy at the end of gravitational collapse, should also be the lower bound of the total entropy of the gauge matter at the end of thermalization in the dual picture, $S_{gauge}\gtrsim S_{AdS}$.  We found that the entropy~(at the mass limit) of the AdS star is an increasing~(decreasing) function of the temperature~(magnetic field), similar behaviour to the mass limit.

The remaining unanswered question is the exact correspondence between the gravitational collapse in the bulk and non-equilibrium deconfinement thermalization of the dual gauge matter.  If bulk gravity is dual to colour-singlet glueball interaction and it causes the gravitational collapse in the AdS, how could the glueball exchange describe the deconfinement thermalization in the dual gauge picture?  Should there exist the critical glueball density corresponding to the mass limit in the bulk which determines the deconfinement phase transition on the gauge theory side?  What is the boundary~(CFT) gauge description of the TOV equation and more generically the Einstein equation in the~(AdS) bulk?  What are the duals of bulk temperature and other thermodynamic and transport quantities such as the adiabatic index and sound speed of the AdS star in the gauge theory side?

\section*{Acknowledgments}

 T.C. would like to thank Sirachak Panpanich for valuable helps.  P.B. is supported in part by the Thailand Research Fund~(TRF) and Commission on Higher Education~(CHE) under grant RMU5380048 and also by the Thailand Center of Excellence in Physics~(ThEP).  T.C. is supported in part by the Thailand Research Fund~(TRF) and Commission on Higher Education~(CHE) under grant RMU5380048.

\appendix

\section{The equation of hydrostatic equilibrium for a spherical symmetric star in $d$ dimensions}\label{appa}

We solve the Einstein equation in $d$-dimensional spacetime in this section.  Starting from the Einstein equation,
    \begin{eqnarray}
        G^{\mu}_{\phantom{1}\nu} = R^{\mu}_{\phantom{1}\nu} - g^{\mu}_{\phantom{1}\nu}\frac{R}{2} = V_{d-2}C_{d-1}T^{\mu}_{\phantom{1}\nu},
    \end{eqnarray}
where $R^{\mu}_{\phantom{1}\nu}$, $g^{\mu}_{\phantom{1}\nu}$, $R$, $T^{\mu}_{\phantom{1}\nu}$, $V_{d-2}$, $C_{d-1}$ are Ricci tensor, metric tensor, Ricci scalar, energy-momentum tensor, the area of $S^{d-2}$ and constant $\left(\frac{16\pi G}{\left(d - 2\right)V_{d-2}c^{4}}\right)$ respectively.  Assuming a perfect fluid, the energy-momentum tensor is given by
    \begin{eqnarray}
        T^{\mu}_{\phantom{1}\nu} =
    \begin{pmatrix}
        \rho c^{2} & & & & \\
        & -P_{r} & & & \\
        & & -P_{\theta_{1}} & & \\
        & & & \ddots & \\
        & & & & -P_{\theta_{d-2}}
    \end{pmatrix},
    \end{eqnarray}
where we use a spherically symmetric metric in $d$ dimensions in the polar coordinates~\cite{Blumenson:1960ma}
    \begin{eqnarray}
        ds^{2} &=& A(r)c^{2}dt^{2} - B(r)dr^{2} - r^{2}d\Omega^{2}_{d-2} \nonumber \\
               &=& A(r)c^{2}dt^{2} - B(r)dr^{2} - r^{2}d\theta^{2}_{1} - r^{2}\sin^{2}{\theta_{1}}\left(d\theta^{2}_{2} + \cdots + \prod_{i=2}^{d-3}\sin^{2}{\theta_{i}d\theta^{2}_{d-2}}\right) \nonumber \\
               &=& A(r)c^{2}dt^{2} - B(r)dr^{2} - r^{2}d\theta^{2}_{1} - r^{2}\sin^{2}{\theta_{1}}\left(d\theta^{2}_{2} + \sum_{j=3}^{d-2}\prod_{i=2}^{j-1}\sin^{2}\theta_{i}d\theta^{2}_{j}\right).
    \end{eqnarray}
The Lagrangian of this metric is then given by
    \begin{eqnarray}
        L = A(r)c^{2}\dot{t}^{2} - B(r)\dot{r}^{2} - r^{2}\dot{\theta}^{2}_{1} - r^{2}\sin^{2}{\theta_{1}}\left(\dot{\theta}^{2}_{2} + \sum_{j=3}^{d-2}\prod_{i=2}^{j-1}\sin^{2}\theta_{i}\dot{\theta}^{2}_{j}\right).
    \end{eqnarray}
We will use the Euler-Lagrange equation to find the equations of motion and read off the connections,
    \begin{eqnarray}
        \partial_{\tau}\left(\frac{\partial L}{\partial\dot{q}}\right) = \frac{\partial L}{\partial q}.
    \end{eqnarray}
Consider $t$ component, the equation of motion is
    \begin{eqnarray}
        \ddot{t} + \frac{A'}{A}\dot{r}\dot{t} = 0,
    \end{eqnarray}
and the connections are
    \begin{eqnarray}
        \Gamma^{t}_{\phantom{1}rt} = \Gamma^{t}_{\phantom{1}tr} = \frac{A'}{2A}.
    \end{eqnarray}
The equation of motion in the $r$ component reads
    \begin{eqnarray}
        \ddot{r} + \frac{A'c^{2}}{2B}\dot{t}^{2} + \frac{B'}{2B}\dot{r}^{2} - \frac{r}{B}\dot{\theta}^{2}_{1} - \frac{r\sin^{2}{\theta}_{1}}{B}\left(\dot{\theta}^{2}_{2} + \sum_{j=3}^{d-2}\prod_{i=2}^{j-1}\sin^{2}{\theta_{i}}\dot{\theta}^{2}_{j}\right) = 0,
    \end{eqnarray}
and the connections in the $r$ component are
    \begin{align}
        \Gamma^{r}_{\phantom{1}tt} = \frac{A'c^{2}}{2B},\Gamma^{r}_{\phantom{1}rr} = \frac{B'}{2B},  & \Gamma^{r}_{\phantom{1}\theta_{1}\theta_{1}} = \frac{-r}{B}, \Gamma^{r}_{\phantom{1}\theta_{2}\theta_{2}} = \frac{-r\sin^{2}\theta_{1}}{B}, \nonumber \\ \ldots, \Gamma^{r}_{\phantom{1}\theta_{j}\theta_{j}} &= \frac{-r\sin^{2}\theta_{1}}{B}\prod_{i=2}^{j-1}\sin^{2}{\theta}_{i}.
    \end{align}
Likewise, the equation of motion in the $\theta_{1}$ component is
    \begin{eqnarray}
        \ddot{\theta_{1}} + \frac{2}{r}\dot{r}\dot{\theta_{1}} - \sin{\theta_{1}}\cos{\theta_{1}}\left(\dot{\theta}^{2}_{2} + \sum_{j=3}^{d-2}\prod_{i=2}^{j-1}\sin^{2}{\theta_{i}}\dot{\theta}^{2}_{j}\right) = 0,
    \end{eqnarray}
and the connections in the $\theta_{1}$ component are
    \begin{align}
        \Gamma^{\theta_{1}}_{\phantom{1}r\theta_{1}} = \Gamma^{\theta_{1}}_{\phantom{1}\theta_{1}r} &= \frac{1}{r}, \Gamma^{\theta_{1}}_{\phantom{1}\theta_{2}\theta_{2}} = -\sin{\theta_{1}}\cos{\theta_{1}}, \nonumber \\ \ldots, \Gamma^{\theta_{1}}_{\phantom{1}\theta_{j}\theta_{j}} &= -\sin{\theta_{1}}\cos{\theta_{1}}\prod_{i=2}^{j-1}\sin^{2}{\theta_{i}},
    \end{align}
where $3\leqslant j \leqslant d-2$.
Similarly, the equation of motion in the $\theta_{2}$ component is
    \begin{eqnarray}
        \ddot{\theta_{2}} + \frac{2}{r}\dot{r}\dot{\theta_{2}} + 2\cot{\theta_{1}}\dot{\theta_{1}}\dot{\theta_{2}} - \sin{\theta_{2}}\cos{\theta_{2}}\sum_{j=4}^{d-2}\prod_{i=3}^{j-1}\sin^{2}{\theta_{i}}\dot{\theta}^{2}_{j}\ = 0,
    \end{eqnarray}
and the relevant connections are
    \begin{align}
        \Gamma^{\theta_{2}}_{\phantom{1}r\theta_{2}} = \Gamma^{\theta_{2}}_{\phantom{1}\theta_{2}r} &= \frac{1}{r}, \Gamma^{\theta_{2}}_{\phantom{1}\theta_{1}\theta_{2}} = \Gamma^{\theta_{2}}_{\phantom{1}\theta_{2}\theta_{1}} = \cot{\theta_{1}}, \nonumber \\ \ldots, \Gamma^{\theta_{2}}_{\phantom{1}\theta_{j}\theta_{j}} &= -\sin{\theta_{2}}\cos{\theta_{2}}\prod_{i=3}^{j-1}\sin^{2}{\theta_{i}},
    \end{align}
where $4\leqslant j \leqslant d-2$.
The equation of motion in the $\theta_{j}~(j \geqslant 3)$ component is
    \begin{align}
        & \ddot{\theta_{j}} + \frac{2}{r}\dot{r}\dot{\theta_{j}} + 2\cot{\theta_{1}}\dot{\theta_{1}}\dot{\theta_{j}} + \frac{2\sum_{l=2}^{j-1}\prod_{\substack{i=2 \\ i\neq l}}^{j-1}\sin{\theta_{l}}\cos{\theta_{l}}\sin^{2}{\theta_{i}}}{\prod_{i=2}^{j-1}\sin^{2}{\theta_{i}}}\dot{\theta_{l}}\dot{\theta_{j}}
        \nonumber \\ & - \sum_{k=j+1}^{d-2}\prod_{i=j+1}^{k-1}\sin{\theta_{j}}\cos{\theta_{j}}\sin^{2}{\theta_{i}}\dot{\theta_{k}^{2}} = 0,
    \end{align}
and the connections in $\theta_{j}$ component are
    \begin{align}
        \Gamma^{\theta_{j}}_{\phantom{1}r\theta_{j}} = \Gamma^{\theta_{j}}_{\phantom{1}\theta_{j}r} = \frac{1}{r}, \Gamma^{\theta_{j}}_{\phantom{1}\theta_{1}\theta_{j}} &= \Gamma^{\theta_{j}}_{\phantom{1}\theta_{j}\theta_{1}} = \cot{\theta_{1}}, \Gamma^{\theta_{j}}_{\phantom{1}\theta_{l}\theta_{j}} = \Gamma^{\theta_{j}}_{\phantom{1}\theta_{j}\theta_{l}} \nonumber \\ = \frac{\prod_{\substack{i=2 \\ i\neq l}}^{j-1}\sin{\theta_{l}}\cos{\theta_{l}}\sin^{2}{\theta_{i}}}{\prod_{i=2}^{j-1}\sin^{2}{\theta_{i}}} = \cot{\theta_{l}}&, \Gamma^{\theta_{j}}_{\phantom{1}\theta_{k}\theta_{k}} = -\sin{\theta_{j}}\cos{\theta_{j}}\prod_{i=j+1}^{k-1}\sin^{2}{\theta_{i}},
    \end{align}
where $2\leqslant l \leqslant j-1$ and $j+1\leqslant k \leqslant d-2$.
The Ricci tensor and Ricci scalar can be calculated from
   \begin{subequations}
   \begin{align}
        R^{\rho}_{\phantom{1}\sigma\mu\nu} =
        \partial_{\mu}{\Gamma^{\rho}_{\phantom{1}\nu\sigma}} - \partial_{\nu}{\Gamma^{\rho}_{\phantom{1}\mu\sigma}} &+ \Gamma^{\rho}_{\phantom{1}\mu\lambda}\Gamma^{\lambda}_{\phantom{1}\nu\sigma} - \Gamma^{\rho}_{\phantom{1}\nu\lambda}\Gamma^{\lambda}_{\phantom{1}\mu\sigma}, \nonumber \\
        R_{\mu\nu} =& R^{\lambda}_{\phantom{1}\mu\lambda\nu}, \nonumber \\
        R = R^{\mu}_{\phantom{1}\mu} &= g^{\mu\nu}R_{\mu\nu}. \nonumber
   \end{align}
   \end{subequations}
After some calculations, we have
    \begin{eqnarray}
        R^{t}_{\phantom{1}t} &=& \frac{A''}{2AB} - \frac{A'B'}{4AB^{2}} - \frac{\left(A'\right)^{2}}{4A^{2}B} + (d-2)\frac{A'}{2rAB}, \nonumber\\
        R^{r}_{\phantom{1}r} &=& \frac{A''}{2AB} - \frac{A'B'}{4AB^{2}} - \frac{\left(A'\right)^{2}}{4A^{2}B} + (d-2)\frac{B'}{2rB^{2}}, \nonumber\\
        R^{\theta_{1}}_{\phantom{1}\theta_{1}} &=& \frac{A'}{2rAB} - \frac{B'}{2rB^{2}} - \frac{(d-3)}{r^{2}}\left(1-\frac{1}{B}\right), \nonumber\\
        R^{\theta_{2}}_{\phantom{1}\theta_{2}} &=& \frac{A'}{2rAB} - \frac{B'}{2rB^{2}} - \frac{(d-3)}{r^{2}}\left(1-\frac{1}{B}\right), \nonumber\\
        R^{\theta_{i}}_{\phantom{1}\theta_{i}} &=& \frac{A'}{2rAB} - \frac{B'}{2rB^{2}} - \frac{(d-3)}{r^{2}}\left(1-\frac{1}{B}\right). \nonumber
    \end{eqnarray}
Consider $G^{t}_{\phantom{1}t} = R^{t}_{\phantom{1}t} - \frac{g^{t}_{\phantom{1}t}}{2}\left(R^{t}_{\phantom{1}t} + R^{r}_{\phantom{1}r} + R^{\theta_{1}}_{\phantom{1}\theta_{1}} + R^{\theta_{2}}_{\phantom{1}\theta_{2}} + \ldots + R^{\theta_{i}}_{\phantom{1}\theta_{i}} + \ldots + R^{\theta_{d-2}}_{\phantom{1}\theta_{d-2}}\right) = V_{d-2}C_{d-1}T^{t}_{\phantom{1}t} \rightarrow R^{t}_{\phantom{1}t} - \left(R^{r}_{\phantom{1}r} + \ldots+R^{\theta_{d-2}}_{\phantom{1}\theta_{d-2}}\right) = 2V_{d-2}C_{d-1}\rho c^{2}$, then
    \begin{eqnarray}
        (d-2)\frac{B'}{rB^{2}} + \frac{(d-2)(d-3)}{r^{2}}\left(1-\frac{1}{B}\right) = 2V_{d-2}C_{d-1}\rho c^{2}, \label{1 st result of Gtt}
    \end{eqnarray}
    \begin{eqnarray}
        B' - \frac{\left(d-3\right)}{r}B = B^{2}\left(\frac{2rV_{d-2}C_{d-1}\rho c^{2}}{\left(d-2\right)} - \frac{\left(d-3\right)}{r}\right). \label{Result of Gtt}
    \end{eqnarray}
If we consider an AdS space(with a negative cosmological constant, $\Lambda$), then the Einstein equation reads
    \begin{eqnarray}
        G^{\mu}_{\phantom{1}\nu} + \Lambda g^{\mu}_{\phantom{1}\nu} = V_{d-2}C_{d-1}T^{\mu}_{\phantom{1}\nu},
    \end{eqnarray}
and equation (\ref{Result of Gtt}) becomes
    \begin{eqnarray}
        B'-\frac{\left(d-3\right)}{r}B = B^{2}\left(\frac{2rV_{d-2}C_{d-1}\rho c^{2}}{\left(d-2\right)} - \frac{\left(d-3\right)}{r} - \frac{2\Lambda r}{\left(d-2\right)}\right).
    \end{eqnarray}
Change $B \rightarrow B^{2}$, so that
    \begin{eqnarray}
        B' - \frac{\left(d-3\right)}{2r}B = B^{3}\left(\frac{r V_{d-2}C_{d-1}\rho c^{2}}{\left(d-2\right)} - \frac{\left(d-3\right)}{2r} - \frac{\Lambda r}{\left(d-2\right)}\right).
    \end{eqnarray}
The solution to this equation is
    \begin{eqnarray}
        B^{2} = \frac{1}{1 - \frac{2c^{2}V_{d-2}C_{d-1}}{\left(d-2\right)r^{d-3}}\int\rho r^{d-2}dr + \frac{2\Lambda r^{2}}{\left(d-2\right)\left(d-1\right)}}.
    \end{eqnarray}
Let $\frac{2\Lambda}{\left(d-2\right)\left(d-1\right)} = \frac{1}{l^{2}}$, then
    \begin{eqnarray}
        B^{2} = \frac{1}{1 - \frac{2c^{2}V_{d-2}C_{d-1}}{\left(d-2\right)r^{d-3}}\int\rho r^{d-2}dr + \frac{r^{2}}{l^{2}}} = \frac{1}{1 - \frac{MC_{d-1}}{r^{d-3}} + \frac{r^{2}}{l^{2}}}. \label{Bsquare in AdS}
    \end{eqnarray}
Also the accumulated mass can be defined to be
    \begin{eqnarray}
        M\left(r\right) = \frac{2V_{d-2}}{\left(d-2\right)}\int\rho r^{d-2}dr. \label{Equation of motion of mass}
    \end{eqnarray}
Consider $G^{r}_{\phantom{1}r} = R^{r}_{\phantom{1}r} - \frac{g^{r}_{\phantom{1}r}}{2}\left(R^{t}_{\phantom{1}t} + R^{r}_{\phantom{1}r} + R^{\theta_{1}}_{\phantom{1}\theta_{1}} + R^{\theta_{2}}_{\phantom{1}\theta_{2}} + \ldots + R^{\theta_{i}}_{\phantom{1}\theta_{i}} + \ldots + R^{\theta_{d-2}}_{\phantom{1}\theta_{d-2}}\right) = V_{d-2}C_{d-1}T^{r}_{\phantom{1}r} \rightarrow R^{r}_{\phantom{1}r} - \left(R^{t}_{\phantom{1}t} + \ldots + R^{\theta_{d-2}}_{\phantom{1}\theta_{d-2}}\right) = 2V_{d-2}C_{d-1}P_{r}$, then
    \begin{eqnarray}
        \frac{\left(d-2\right)A'}{rAB} - \frac{\left(d-2\right)\left(d-3\right)}{r^{2}}\left(1-\frac{1}{B}\right) = 2V_{d-2}C_{d-1}P_{r}. \nonumber
    \end{eqnarray}
Use equation (\ref{1 st result of Gtt}) from $G^{t}_{\phantom{1}t}$ and multiply by $rB/(d-2)$,
    \begin{eqnarray}
        \frac{A'}{A} + \frac{B'}{B} = \frac{2V_{d-2}C_{d-1}}{\left(d-2\right)}r B\left(\rho c^{2} + P_{r}\right). \label{Gtt Grr 2}
    \end{eqnarray}
Change $A \rightarrow A^{2}$, $B \rightarrow B^{2}$, equation (\ref{Gtt Grr 2}) becomes
    \begin{eqnarray}
        \frac{A'}{A} + \frac{B'}{B} = \frac{V_{d-2}C_{d-1}rB^{2}}{\left(d-2\right)}\left(\rho c^{2} + P_{r}\right), \nonumber
    \end{eqnarray}
Solve this equation to find relations between $A$ and $B$,
    \begin{eqnarray}
        A^{2}\left(r\right) = \frac{e^{2\chi\left(r\right)}}{B^{2}\left(r\right)},
    \end{eqnarray}
where
    \begin{eqnarray}
        \chi\left(r\right) = \frac{V_{d-2}C_{d-1}}{\left(d-2\right)}\int\left(\rho\left(r\right)c^{2} + P_{r}\left(r\right)\right)rB^{2}\left(r\right)dr. \label{Equation of motion of chi}
    \end{eqnarray}
Finally we obtain the coupled equations of motion from equation (\ref{Equation of motion of mass}) and (\ref{Equation of motion of chi})
    \begin{subequations}
    \begin{align}
        M'\left(r\right) &= \frac{2V_{d-2}}{\left(d-2\right)}\rho\left(r\right)r^{d-2}, \\ \chi'\left(r\right) &= \frac{V_{d-2}C_{d-1}}{\left(d-2\right)}\left(\rho\left(r\right)c^{2} + P_{r}\left(r\right)\right)rB^{2}\left(r\right).
    \end{align}
    \end{subequations}
Moreover, when we consider the energy momentum conservation $\nabla_{\mu}T^{\mu}_{\phantom{1}\nu} = 0$ by letting $\nu = r$ and $P_{r} = P_{\theta_{1}} = \ldots = P_{\theta_{i}} = \ldots = P_{\theta_{d-2}} = P$, $A \rightarrow A^{2}$, it leads to the TOV equation in $d$-dimension,
    \begin{eqnarray}
        \frac{dP}{dr} = -\left(\rho c^{2} + P\right)\frac{A'}{A}. \nonumber
    \end{eqnarray}
Next we want to rewrite this equation in the form containing thermodynamic quantities such as the chemical potential, the entropy, and the temperature of matter within the spherically symmetric star.  From thermodynamic relations involving the entropy density $s$;
\begin{eqnarray}
sT & = & P + \rho c^{2} - \mu n,  \\
s~dT & = & dP - n~d\mu,
\end{eqnarray}
the TOV equation can be rewritten as
\begin{equation}
s\left(T' + T \frac{A'}{A}\right)+n\left(\mu'+\mu \frac{A'}{A}\right) = 0,
\end{equation}
implying two equations to be satisfied simultaneously
\begin{eqnarray}
T' + T \frac{A'}{A} = \mu'+\mu \frac{A'}{A} = 0.
\end{eqnarray}
The temperature equation can be solved to obtain the redshifted temperature profile within the star $T=T_{0}/A(r)$ where $A(0)=1$ and $T_{0}$ is the temperature at the star center.  The chemical potential equation similarly gives
    \begin{eqnarray}
        \mu(r) = \frac{\mu_{0}}{A(r)}.
    \end{eqnarray}
The coupled equations of motion can then be written in terms of the accumulated mass and chemical potential as the following
    \begin{subequations}
    \begin{align}
        M'\left(r\right) &= \frac{2V_{d-2}}{\left(d-2\right)}\rho\left(r\right)r^{d-2}, \label{EoM of Mapp} \\ \mu'\left(r\right) &= \mu\left(r\right)\left(\frac{B'(r)}{B(r)} - \frac{V_{d-2}C_{d-1}}{\left(d-2\right)}\left(\rho\left(r\right)c^{2}+ P_{r}\left(r\right)\right)rB^{2}\left(r\right)\right). \label{EoM of muapp}
    \end{align}
    \end{subequations}

\section{Euler-Maclaurin Formula} \label{appb}

A slowly converging series can be evaluated effectively by using an integral as in the Euler-Maclaurin formula
\begin{eqnarray}
\sum_{j=0}^{\infty}f\left(j+\frac{1}{2}\right) \approx \int_{0}^{\infty}~f(x)~dx + \frac{1}{24}(f'(0)-f'(\infty))+O(x^{3}).
\end{eqnarray}
In this article, the partition function sum over Landau states is approximated using this conventional method by letting
\begin{eqnarray}
f(x) & = & \ln \left( 1 + \exp{\left(\frac{\mu-\sqrt{m^2c^4+p_{n}^{2}c^{2}+4xmc^{2}\mu_{B}B}}{k_{B}T}\right)} \right),
\end{eqnarray}
where $x=j+1/2$.

\section{Dimensional translation table}\label{appc}
    \begin{tabular}{ | c | c | c | }
        \hline
        quantity & dimensionless variable & physical variable \\ \hline
        density & $\rho$ & $\rho_{0}\rho$ \\ \hline
        pressure & $P$ & $\rho_{0}P$ \\ \hline
        mass & $M$ & $\left(\frac{c^{10}}{G^{4}\rho_{0}}\right)^{\frac{1}{3}}M$ \\ \hline
        radius & $r$ & $\left(\frac{c^{4}}{G\rho_{0}}\right)^{\frac{1}{3}}r$ \\ \hline
        temperature & $T$ & $\frac{\left(\rho_{0}c^{4}\hbar^{4}\right)^{\frac{1}{5}}}{k_{B}}T$ \\ \hline
        magnetic field & $B$ & $\frac{\left(\rho_{0}c^{4}\hbar^{4}\right)^{\frac{1}{5}}}{\mu_{B}}B$ \\
        \hline
    \end{tabular}
$\rho_{0} = \frac{\left(\frac{m_{p}}{m_{s}}c^{2}\right)^{5}}{c^{4}\hbar^{4}}$ where $m_{p}$ and $m_{s}$ are the rest mass of particles and the mass used in simulation, respectively.

\end{document}